\documentclass[prd,aps,graphics]{revtex4}
\usepackage{epsfig} 
\DeclareGraphicsExtensions{.eps,.ps,.eps.gz,.ps.gz,.eps.Z,{}}

\newcommand{\be}{\begin{equation}} 
\newcommand{\ee}{\end{equation}} 
\newcommand{\bea}{\begin{eqnarray}} 
\newcommand{\eea}{\end{eqnarray}}

\newcommand{\Gaunt}[1]{{}^{(#1)}\,\!{\cal G}}
\newcommand{\ttl}{{}^{\rm line}\,\!\tilde{t}}
\newcommand{\tts}{{}^{\rm star}\,\!\tilde{t}}

\def\vx{\vec{x}} 
\def\vk{\vec{k}}

\def\vl{\vec{\ell}} 
 
\def\hk{\hat{k}} 
\def\hg{\hat{\gamma}} 
 
\def\rs{r \! _{\ast}}

\def\dd{ {\rm d} } 
\def\vdg{\delta\vec{\gamma} }
\def\hT{\frac{\hat{\delta T}}{T} }
\def\Phib{\overline{\Phi}}
\def\mF{{\cal F}}
\def\mg{\langle}
\def\md{\rangle}
\def\nul{\nu_{3}^{\rm line}}
\def\nus{\nu_{3}^{\rm star}}
 
\begin{document} 
\title{The shape of high order correlation functions in CMB anisotropy maps} 
 
\author{Tristan Brunier}
\email{Tristan.Brunier@cea.fr} 
\affiliation{Service de Physique Th{\'e}orique, 
         CEA/DSM/SPhT, Unit{\'e} de recherche associ{\'e}e au CNRS, CEA/Saclay 
         91191 Gif-sur-Yvette c{\'e}dex, France.}

\author{Francis Bernardeau}
\email{Francis.Bernardeau@cea.fr} 
\affiliation{Service de Physique Th{\'e}orique, 
         CEA/DSM/SPhT, Unit{\'e} de recherche associ{\'e}e au CNRS, CEA/Saclay 
         91191 Gif-sur-Yvette c{\'e}dex, France.}

\vskip 0.15cm

\begin{abstract}
We present a phenomenological investigation of non-Gaussian effects that could be seen on CMB temperature maps. Explicit expressions for the temperature correlation functions are given for different types of primordial mode couplings. We argue that a simplified description of the radial transfer function for the temperature anisotropies allows to get insights into the general properties of the bi and tri-spectra. The accuracy of these results is explored together with the use of the small scale approximation to get explicit expressions of high order spectra. The bi-spectrum is found to have alternate signs for the successive acoustic peaks. Sign patterns for the trispectra are more complicated and depend specifically on the type of metric couplings. Local primordial couplings are found to give patterns that are different from those expected from weak lensing effects.
\end{abstract} 
 
 \vskip2pc

\maketitle 

\section{Introduction}

The standard inflationary models~\cite{inflation} have been very 
successful in explaining the basic features of the CMB observations: 
near scale invariant power spectrum of, to a good approximation, 
primordial Gaussian adiabatic  perturbations. These are undoubtedly 
important results that give support to the inflationary scheme. From 
another point of view it is an uncomfortable situation since there is 
practically no possibility  with the current data sets to distinguish 
different models of inflations.
The search of observational signatures beyond predictions from generic 
inflation is therefore crucial for getting insights into the nature of 
the inflaton field or more generally into physics of the early 
Universe. For instance trans-Planckian effects~\cite{Transplanck}, 
presence of isocurvature modes~\cite{Iso1, Iso2, curvaton} and 
non-Gaussian effects~\cite{bartoloetal} could betray some aspects of 
the inflationary physics. It is now well established that in standard 
inflation, e.g. single field inflation with slow roll conditions, a 
minimal amount of non-Gaussianities is expected to be induced along the 
cosmic evolution~\cite{bartoloetal}. They simply come from the 
couplings of modes contained in the Einstein equations, that are 
intrinsically nonlinear in the fields. The more famous of those are the 
lensing effects on the CMB sky~\cite{CMBLens,TriLens, HuLens}. They 
indeed can be seen as the effects of couplings between the 
gravitational potential on the last-scattering  surface and those 
present on the line-of-sights. Of course other local couplings exist 
that take place before or during the recombination period. For instance 
Bartolo and collaborators in  \cite{bartoloetal} derive the amplitude 
and nature of the superhorizon couplings of the gravitational 
potentials in terms of their expected bi-spectrum. Its shape betrays 
the nature and shape of the quadratic couplings in the Einstein 
equations that generically are expected to determine the types of 
couplings for the modes that reenter the horizon.

For one-field inflation couplings at horizon crossing are rather 
generic and induce typically $10^{-5}$ effects in the metric 
perturbations. Effects of such amplitude are expected to be marginally 
visible. They are however expected to be altered by second order 
effects taking place at low redshift after horizon crossing. Except for 
the Sachs Wolfe plateau or the lens effects, predictions of the 
observable intrinsic non-Gaussian effects are not known. They require 
the computation of the physics of recombination up to second order. An 
enterprise still to be done.

There are however cases where significant mode couplings could survive 
the inflationary period. This is the case for some flavors of the 
curvaton models~\cite{curvaton}. This could also be the case in other 
models of multi-field inflation where a transfer of modes, from 
isocurvature to adiabatic, is possible during or at the end of the 
inflationary period as described in \cite{BU1, BU2, BKU}. In this case 
no isocurvature modes are expected to survive, contrary to the case of 
the curvaton model, and the models
predict only adiabatic fluctuations the spectrum of which can be 
arbitrarily close to scale invariance. However nothing prevents the 
initial isocurvature modes from developping significant non-Gaussian 
features which may be transfered into those of the adiabatic 
fluctuations. Such a transfer mechanism has been described in details 
in \cite{BU3} where the resulting high order correlation properties are 
explicitly given.

In principle, the properties of the metric fluctuations entirely 
determine those of the temperature or polarization CMB sky. However, 
the transcription of the properties of metric
to properties in the observed temperature field reveals involved for 
at least two reasons. First, it is to be noted that the transfer 
physics is naturally  best encoded in Fourier space where all modes 
evolve independently from one-another whereas the nonlinear couplings 
of Eq.~(\ref{Ftransform}) is local in real space. The second reason is 
that projection effects should be taken into account so that the modes 
that are observed correspond to a collection of Fourier modes. The then 
simple functional forms (such those of Eqs.~(\ref{Bispec}) 
and~(\ref{Trispec}) for instance) that are rather generically expected 
for the metric are then greatly altered. For instance the sign of the 
three- and four-point correlation function may depend on scale. The aim 
of the investigations pursued in this paper is to try to uncover simple 
prescriptions for the shape, e.g. angular dependence, of those 
quantities. Comparison with the trispectra induced by weak lensing 
effects will also be presented.

The paper is divided as follows. In the second section we detail the 
couplings we assume for the primordial metric fluctuations.
  In the third section we present the basic quantities that are required 
to describe the way the potential fluctuations are transfered into the 
temperature fluctuations. We then take advantage of these results to 
present the theoretical shapes of the bi- and tri-spectra of the 
temperature field in case for the models of potential high-order 
correlation functions described in the previous paragraph. The 
following section is devoted to an attempt to obtain a phenomenological 
description of the transfer function that gives insights into the 
angular dependence of these correlation functions. The last section is 
devoted to computation in the small angle approximation. Calculations 
are obviously much more straightforward in this limit which indeed 
corresponds to a regime where most of the observations can be made.

\section{A model of primordial non-Gaussian metric perturbations}

In the analysis pursued in this paper we assume that the primordial 
metric fluctuations are those expected in rather generic models of 
multiple-field inflation, more specifically along the descriptions 
presented in \cite{BU1,BU2,FiniteVolBU}. In this family of models, the 
surviving couplings in the metric, are expected to be, to a good 
approximation, equivalent to those induced by the superposition of two 
stochastically independent fields, a Gaussian one and one obtained by a 
non-linear transform of a Gaussian field with the same spectrum. In 
other words the local potential would read,
\begin{equation}
\Phi(\vx)=\cos \alpha\ \Phi_{1}(\vx)+ \sin \alpha\ \mF[\Phi_{2}(\vx)]
\end{equation}
where ratio between the initial adiabatic and isocurvature fluctuations 
is described by the mixing angle $\alpha$ and where the function $\mF$ 
takes into account the self coupling of the field that gave rise to 
this part of the metric fluctuations. The function $\mF$ obviously 
depends on the details of the inflationary model in particular on the 
self interaction potential along the transverse directions. As argued 
in \cite{BU1} a natural choice is a quartic potential. In such a case, 
the function $\mF$ is characterized by two quantities. One is the 
amplitude of the coupling constant times $N_{e}$, the number of 
e-foldings between horizon crossing of the observable modes and their 
transfer from isocurvature to adiabatic direction, the other is related 
to the field value in the transverse direction $\Phib$ at our current 
Hubble scale. The latter actually corresponds to a finite volume effect 
\cite{FiniteVolBU}. In this framework the function
$\mF$ then reads\footnote{In~\cite{BU1}, the nonlinear transform 
function was derived for the transverse field fluctuations. In this 
case the parameter $\mu_{3}$ in (\ref{Ftransform}) is $\mu_{3}=-\lambda 
N_{e}/(3H^2)$. A  similar transform applies to the induced metric 
fluctuations since the two are proportional  to each other with a 
redefinition of the coupling parameter $\mu_{3}$.}
\begin{equation}
\label{Ftransform}
\mF(\Phi)=\frac{\Phi+\Phib}{\sqrt{1-\mu_{3}(\Phi+\Phib)^2/3}},
\end{equation}
where $\mu_{3}$ is related to the self coupling of the field. For a 
coupling in $\lambda\phi^4/4$ we have
\begin{equation}
\mu_{3}\sim-\lambda N_{e}/C_{2}^2\end{equation}
where $C_{2}$ is the amplitude of the metric fluctuation at 
super-Hubble scale, $C_{2}\approx 10^{-5}$.
One important feature of this description is that the non linear 
transform of the field is local in real space. This description has the 
advantage of providing a full description of the model. This equation 
should however be used with care. It gives a good account of the mode 
couplings only when the effective coupling constant $\lambda N_{e}$ is 
small. It is therefore preferable to consider the bi- and tri-spectra 
of the potential field. So let us define the spectrum $P(k)$, bi- and 
tri-spectra, $B(\vk_{1},\vk_{2})$ and  $T(\vk_{1},\vk_{2},\vk_{3})$, of 
the potential field the following way,
\begin{eqnarray}
\mg \Phi_{\vk_{1}} \Phi_{\vk_{2}}\md&=&\delta_{\rm 
Dirac}(\vk_{1}+\vk_{2})\ P(k_{1})\\
\mg \Phi_{\vk_{1}} \Phi_{\vk_{2}} \Phi_{\vk_{3}}\md&=&\delta_{\rm 
Dirac}(\vk_{1}+\vk_{2}+{\vk_{3}})\ 
B(\vk_{1},\vk_{2},\vk_{3})\label{BispecFormal}\\
\mg \Phi_{\vk_{1}} \dots \Phi_{\vk_{4}}\md_{c}&=&\delta_{\rm 
Dirac}(\vk_{1}+\dots+{\vk_{4}})\ T(\vk_{1},\vk_{2},\vk_{3},\vk_{4}),
\label{TrispecFormal}\end{eqnarray}
where the underscript $_{c}$ stands for the connected part of the 
corresponding ensemble average.
As $\Phi_{1}$ and $\Phi_{2}$ have the same power spectrum, if the 
nonlinear couplings of $\Phi_{2}$ are small then the power spectrum of 
the metric fluctuations is left unchanged
(to corrections that are quadratic in the coupling parameter). Then for 
the form presented in Eq. (\ref{Ftransform}), for a small coupling 
parameter $\mu_{3}$ we have,
\begin{equation}
\label{Bispec}
B(\vk_{1},\vk_{2},\vk_{3})=\nu_{2}\left[P(k_{1})P(k_{2})+{\rm 
perm.}\right]
\end{equation}
with $\nu_{2}=\sin^3\alpha\,\mu_{3}\Phib$, $\Phib$ being a priori of 
the order of $H$ being the value of the Hubble constant at horizon 
crossing, and
\begin{equation}
\label{Trispec}
T(\vk_{1},\vk_{2},\vk_{3})=\nus\left[P(k_{1})P(k_{2})P(k_{3})+{\rm 
perm.}\right]+
\nul\left[P(k_{1})P(k_{2})P(\vert \vk_{2}+\vk_{3}\vert)+{\rm 
perm.}\right]
\end{equation}
with $\nus=\sin^4\alpha\,\mu_{3}$ and 
$\nul=\sin^4\alpha\,\mu_{3}^2\Phib^2$.

Note that if we are interested in the bispectrum, the parameter 
$\nu_{2}$ is identical to the parameter $f_{NL}$ usually used to 
describe the nonlinear transform of the metric 
fluctuations~\cite{TheseKomatsu}. It is to be noted however that 
whereas $f_{NL}$ is expected of the order unity in single field 
inflation, $\nu_{2}$ can reach much larger values in multiple-field 
inflation. This is one justification of the investigations presented in 
this paper.

We can see that the tri-spectrum contains two terms of different 
geometrical shapes. The relative importance of these two terms depends 
in particular on the value of $\Phib$. This ratio is naturally of order 
unity, but otherwise totally unpredictable~\footnote{At best what a 
complete theory would give is the expected probability distribution 
function of $\Phib$, a computation sketched in \cite{FiniteVolBU}.}. In 
the following we therefore consider the two cases and examine the 
consequences of both terms on the temperature high order correlation 
functions.

\section{Expressions of the correlation functions} 

In this section we set the basic relations that relate the primordial metric fluctuations to the temperature fluctuations.
 
\subsection{Radial transfer function}
 
The observed temperature anisotropies are decomposed in a sum of spherical harmonics with coefficients $a_{\ell m}$. The inverse relation gives the expression of those coefficients as a function of the temperature fluctuation  $\frac{\delta T}{T}(\hg)$ in the direction $\hg$ on the sky, 
\be 
a_{\ell m} = \int {\dd}^2 \hg \frac{\delta T}{T}(\hg) Y^\ast_{\ell m}(\hg).
\ee 
When the linear theory is applied to the metric and density fluctuations the coefficient $a_{\ell m}$ are linearly related to the primordial metric fluctuations. In the following we will only take into account the existence of adiabatic fluctuations. Then one can write,
\be 
\label{alm} 
a_{\ell m} = 4 \pi (-i)^{\ell} \int \frac{{\rm d}^3 \vk}{(2 \pi)^{3/2}} T_{\ell}(k) \Phi(\vk) Y^\ast_{\ell m}(\hk), 
\ee 
where $\Phi(\vk)$ is the three-dimensional Fourier transformed of the gravitational potential, $T_{\ell} (k)$ is the photon transfer function and $Y_{\ell m}(\hat{k})$ are the spherical harmonics in the direction given by the unit vector $\hk=\vk / k $. 
 
The functions $T_{\ell} (k)$ encode all the micro-physics that takes place after horizon crossing until the photons reach the observer. There is one well known limit case which corresponds to large-angular scales for vanishing curvature and cosmological constant. In this limit indeed, the observed local temperature fluctuation is simply one third of the local potential. This is the so-called Sachs-Wolfe effect \cite{SachsWolfe}. In this case we simply have  $T_{\ell} (k)=-1/3 j_{\ell}(k r_{\ast})$ neglecting the late and early Integrated Sachs Wolfe effects. Although the validity regime of this form is limited, it is worth investigating since it corresponds to a case where only projections effects have to be taken into account.

One aim of this paper is to take advantage of the expression (\ref{alm}) to relate as accurately as possible the statistical properties of the $a_{\ell m}$ to those of the potential field. To do so we found fruitful to introduce the radial transfer function. Let us first decompose the gravitational potential intercepted on a sphere of radius $r$ as 
\be 
\Phi(r \hg)= \sum_{\ell m} \Phi_{\ell m}(r) Y_{\ell m}(\hg). 
\ee 
Note that conversely 
\be 
\Phi_{\ell m}(r) = \int {\rm d}^2 \hg \Phi(r \hg) Y^\ast_{\ell m}(\hg). 
\ee 
In Fourier space, we then get 
\bea 
\Phi(\vk) &=& \int \frac{ {\rm d}^3 \vx }{ (2 \pi)^{3/2} } \Phi(\vx) {\rm e}^{i \vk \cdot \vx} \;, \\ 
&=&\frac{1}{(2 \pi)^{3/2}} \sum_{\ell m} \int   r^2 {\rm d}r \; {\rm d}^2 \hg \; \Phi _{\ell m}(r) Y_{\ell m}(\hg) {\rm e}^{i \vk \cdot \vx}. 
\eea 
Using the decomposition of plane waves on a sphere 
\be 
\label{expbes} 
{\rm e}^{i \vk \cdot \vx} = 4 \pi \sum_{LM} i^L j_L(k r) Y_{LM}(\hk) Y^\ast_{LM}(\hg), 
\ee 
with $\vec{x}=r \hg$, one obtains
\bea 
\Phi(\vk) &=& 4 \pi \sum_{L,l,M,m} i^L \int \frac{ {\rm d}r }{ (2 \pi)^{3/2} } r^2 \int {\rm d}^2 \hg  j_L(k r) Y_{LM}(\hk) Y^\ast_{LM}(\hg) \Phi _{\ell m}(r) Y_{\ell m}(\hg),  \\ 
&=& \sqrt{ \frac{2}{\pi} } \sum_{\ell m} i^{\ell} \int {\rm d}r r^2 j_{\ell}(k r) Y_{\ell m}(\hk) \Phi _{\ell m}(r). 
\eea 
Conversely 
\be 
\Phi_{\ell m}(r)=(-i)^{\ell} \sqrt{ \frac{2}{\pi} } \int {\rm d}^3 k \Phi(\vk) j_{\ell}(k r) Y^\ast_{\ell m}(\hk). 
\ee 
Hence Eq. (\ref{alm}) can be recast in 
\be 
a_{\ell m} = \frac{ 2}{ \pi } \int {\rm d}r R_{\ell}(r) \Phi_{\ell m}(r), 
\ee 
where the radial transfer function $R_{\ell}(r)$ is obtained through a radial Fourier transform of the initial Fourier space transfer functions, 
\be 
\label{radexpress} 
R_{\ell}(r)= \int{\rm d} k  k^2 r^2 T_{\ell}(k) j_{\ell}(k r).  
\ee 
Obviously the micro-physics of the recombination can equally be described by the functions $T_{\ell}(k)$ or by the functions $R_{\ell}(r)$. However the latter are more directly related to the physics of recombination, which takes place over a short range of radius, and one can therefore try to find approximate forms for their $r$ dependence. The Sachs Wolfe regime corresponds to a limit case for which,
 \begin{equation}
 R^{\rm SW}_{\ell}(r)= -\frac{\pi}{6}\delta_{\rm Dirac}(r-r_{\ast}).
 \end{equation} 

Before we propose approximate forms for $R_{\ell}(r)$ let us first detail how the statistical quantities we are interested in are related to this function.
 
\subsection{Power spectrum} 
 
We first want to express the ensemble average of  products of $a_{\ell m} a^*_{\ell' m'}$ as a function of the potential power spectrum and the radial transfer function.  Using Eq.~(\ref{radexpress}) we naturally get 
\bea
\langle a_{\ell m} a^*_{\ell' m'} \rangle &=& \delta_{\ell \ell'} \delta_{m m'} \left( \frac{ 2}{ \pi} \right)^3 \int {\rm d}r  \, r^2  \int {\rm d}r' \, r'^2 \int {\rm d}k \, k^2 P(k) j_{\ell}(kr) j_{\ell'}(kr') \nonumber\\
&&\int {\rm d}k_1   \, k_1^2 T_{\ell}(k_1) j_{\ell}(k_1 r) \int {\rm d}k_2 \,  k_2^2  T_{\ell'}(k_2) j_{\ell'}(k_2 r'), 
\eea
where the integration over $r'$ may be performed if one takes advantage of  
\be 
\label{dirac} 
\int {\rm d}x \; x^2 j_{\ell}(k x) j_{\ell}(k' x) = \frac{ \pi }{ 2 k^2 } \delta_{\rm Dirac}(k-k'). 
\ee 
Then temperature anisotropy power spectrum $C_{\ell}$, defined as,
\be 
\langle a_{\ell m} a^*_{\ell' m'} \rangle = (-1)^{m'} \langle a_{\ell m} a_{\ell' -m'} \rangle = \delta_{\ell \ell'} \delta_{m m'}  C_{\ell} \;, 
\ee  
reads
\be 
\label{Clexpr} 
 C_{\ell}  = \left( \frac{ 2}{ \pi} \right)^2 \int {\rm d}r R_{\ell}(r) \xi_{\ell}(r) \;, 
\ee 
where we define 
\be 
\label{xiexpr} 
\xi_{\ell}(r) = \int {\rm d}k k^2 P(k) T_{\ell}(k) j_{\ell}(k r).  
\ee 
We will see now that the functional relation (\ref{Clexpr}) can be generalized to higher order correlation functions.

\subsection{Bispectrum} 
 
\subsubsection{Full sky expression}

We are interested in the ensemble average of the product of three $a_{\ell m}$ that should not vanish when the potential field exhibits non-Gaussian properties. This correlation function formally reads,
\be 
\label{alm3} 
\langle a_{\ell_1 m_1} a_{\ell_2 m_2} a_{\ell_3 m_3} \rangle = \left( \frac{ 2}{ \pi} \right)^3 \int {\rm d}r_1 \int {\rm d}r_2 \int {\rm d}r_3 R_{\ell_1}(r_1) R_{\ell_2}(r_2) R_{\ell_3}(r_3) \langle \Phi_{\ell_1 m_1}(r_1) \Phi_{\ell_2 m_2}(r_2) \Phi_{\ell_3 m_3}(r_3) \rangle \;. 
\ee 
The ensemble average that appears in the r.h.s. of this equation can be related to the potential bispectrum,
\bea 
\label{generalphi3} 
\langle \Phi_{\ell_1 m_1}(r_1) \Phi_{\ell_2 m_2} (r_2) \Phi_{\ell_3 m_3} (r_3) \rangle &=& (-i)^{\ell_1+\ell_2+\ell_3} \left( \frac{2}{\pi} \right)^{3/2} \int {\rm d}^3 \vk_1 \; {\rm d}^3 \vk_2 \; {\rm d}^3 \vk_3 \; \langle \Phi(\vk_1) \Phi(\vk_2) \Phi(\vk_3) \rangle \nonumber\\ 
&& j_{\ell_1}(k_1 r_1) j_{\ell_2}(k_2 r_2) j_{\ell_3}(k_3 r_3) \; Y_{\ell_1 m_1}^*(\hk_1) Y_{\ell_2 m_2}^*(\hk_2) Y_{\ell_3 m_3}^*(\hk_3) \; . 
\eea 
The Dirac distribution that appears in (\ref{BispecFormal})
can be rewritten as a Fourier transform
\be 
\label{diracexp} 
\delta_{\rm Dirac} \left( \vk_1+\vk_2+\vk_3 \right) = \int \frac{ {\rm d}^3 \vx }{(2 \pi)^{3} } {\rm e}^{i (\vk_1+\vk_2+\vk_3) \cdot \vx}\;, 
\ee 
and then expanded into spherical Bessel functions with Eq.~(\ref{expbes}) 
\bea 
\label{diracbes} 
\delta_{\rm Dirac} \left( \vk_1+\vk_2+\vk_3 \right) &=& (4\pi)^3 \int \frac{ {\rm d}^3 \vx }{(2 \pi)^{3} } \sum_{L_1,L_2,L_3,M_1,M_2,M_3} i^{L_1+L_2+L_3} j_{L_1}(k_1 x) j_{L_2}(k_2 x) j_{L_3}(k_3 x) \nonumber\\ 
&& Y_{L_1 M_1}(\hk_1) Y_{L_2 M_2}(\hk_2) Y_{L_3 M_3}(\hk_3) Y_{L_1 M_1}^*(\hg) Y_{L_2 M_2}^*(\hg) Y_{L_3 M_3}^*(\hg), 
\eea 
where $\vec{x}=x \hg$.
Using the form (\ref{Bispec}) for the expression of the bispectrum and inserting Eq.~(\ref{diracbes}) into (\ref{generalphi3}), we get 
\bea 
\label{generalphi3bis} 
\langle \Phi_{\ell_1 m_1}(r_1) \Phi_{\ell_2 m_2} (r_2) \Phi_{\ell_3 m_3} (r_3) \rangle_c &=& 8 \nu_2 \left( \frac{2}{\pi} \right)^{3/2} \Gaunt{3}_{\ell_1 \, \ell_2 \, \ell_3}^{m_1  m_2 m_3} \int  {\rm d}x \; x^2 \int {\rm d} k_1 \;  k^2_1 j_{\ell_1}(k_1 r_1) j_{\ell_1}(k_1 x) \nonumber \\ 
&& \int {\rm d} k_2  \; k^2_2  j_{\ell_2}(k_2 r_2) j_{\ell_2}(k_2 x) \int {\rm d} k_3 \; k^2_3 j_{\ell_3}(k_3 r_3) j_{\ell_3}(k_3 x)  \nonumber\\ 
&&  \left[ P(k_1) P(k_2) + P(k_2) P(k_3) + P(k_3) P(k_1) \right] \;, 
\eea 
where the Gaunt integral is defined by 
\be 
\label{Gaunt3}
\Gaunt{3}_{\ell_1 \, \ell_2 \, \ell_3}^{m_1  m_2 m_3}= \int {\rm d}^2 \hg \; Y_{\ell_1 m_1}(\hg) Y_{\ell_2 m_2}(\hg) Y_{\ell_3 m_3}(\hg) \; .  
\ee  
We integrate over the momenta $k_i$ which do not appear as an argument $P(k_i)$ in Eq.~(\ref{generalphi3bis}) and then perform the integrations over two of the radial variables. Eq.~(\ref{alm3}) then becomes 
\bea 
\label{bisexpr} 
\langle a_{\ell_1 m_1} a_{\ell_2 m_2} a_{\ell_3 m_3} \rangle_c &=& 8 \nu_2 \left( \frac{ 2}{ \pi} \right)^{3 / 2}  \Gaunt{3}_{\ell_1 \, \ell_2 \, \ell_3}^{m_1  m_2 m_3} \nonumber\\ 
&&\int {\rm d}r \left[ R_{\ell_1}(r) \xi_{\ell_2}(r) \xi_{\ell_3}(r) +  R_{\ell_2}(r) \xi_{\ell_3}(r) \xi_{\ell_1}(r) + R_{\ell_3}(r) \xi_{\ell_1}(r) \xi_{\ell_2}(r) \right] \; . 
\eea 
 
The bispectrum is expressed in terms of the fundamental functions $R_\ell$ and $\xi_\ell$ that appeared in the computation of the power spectrum. We shall see a diagrammatic interpretation of each of these terms in section~\ref{sec:diagram}. For instance, the third term of Eq.~(\ref{bisexpr}) is diagrammatically represented in Fig.~\ref{bispectre}.

The expression~(\ref{bisexpr}) also involves geometrical factors encoded by the Gaunt integrals. In the following, we introduce an estimator which should allow to define a {\em reduced} quantity. We also present the correspondence with the small angle approximated bispectrum. 

\subsubsection{Estimator}

Following~\cite{TheseKomatsu, BestGangui} an unbiased estimator of the angular averaged bispectrum may be chosen to be
\be
B_{\ell_1 \ell_2 \ell_3} = \sum_{m_1,m_2,m_3} \left( \begin{array}{crcl} \ell_1 & \ell_2 & \ell_3 \\ m_1 & m_2 & m_3  \end{array}\right) a_{\ell_1 m_1} a_{\ell_2 m_2} a_{\ell_3 m_3} \;, 
\ee
where the Wigner 3-j symbol is related to the Gaunt integral by
\be
\Gaunt{3}_{\ell_1 \, \ell_2 \, \ell_3}^{m_1  m_2 m_3} = \sqrt{ \frac{(2 \ell_1+1)(2 \ell_2 +1)(2\ell_3 +1)}{4 \pi} }  \left( \begin{array}{crcl} \ell_1 & \ell_2 & \ell_3 \\ m_1 & m_2 & m_3  \end{array}\right)  \left( \begin{array}{crcl} \ell_1 & \ell_2 & \ell_3 \\ 0 & 0 & 0  \end{array}\right) \; .
\ee
Using the equality
\be
\sum_{m_1,m_2}  \left( \begin{array}{crcl} \ell_1 & \ell_2 & \ell_3 \\ m_1 & m_2 & m_3  \end{array}\right)  \left( \begin{array}{crcl} \ell_1 & \ell_2 & \ell'_3 \\ m_1 & m_2 & m'_3  \end{array}\right) = \frac{\delta_{\ell_3, \ell_3'} \delta_{m_3, m_3'}}{2\ell_3+1}\;, 
\ee
we get
\be
\langle B_{\ell_1 \ell_2 \ell_3} \rangle_c =  \sqrt{ \frac{(2 \ell_1+1)(2 \ell_2 +1)(2\ell_3 +1)}{4 \pi} }  \left( \begin{array}{crcl} \ell_1 & \ell_2 & \ell_3 \\ 0 & 0 & 0  \end{array}\right) b_{\ell_1 \ell_2 \ell_3} \;, 
\ee
with the reduced bispectrum defined as
\be
\label{fullbis}
b_{\ell_1 \ell_2 \ell_3} = 8 \nu_2 \left( \frac{2}{\pi} \right)^{3/2} \int {\rm d}r \left[ R_{\ell_1}(r) \xi_{\ell_2}(r) \xi_{\ell_3}(r) +  R_{\ell_2}(r) \xi_{\ell_3}(r) \xi_{\ell_1}(r) + R_{\ell_3}(r) \xi_{\ell_1}(r) \xi_{\ell_2}(r) \right] \; .
\ee
 The reduced bispectrum reveals convenient to describe the non-Gaussian part of the signal as it does not include the overall geometrical factors.

It is also convenient to introduce a normalized bispectrum $\tilde{b}_{\ell_1,\ell_2,\ell_3}$ defined by
\be
\label{Normbis}
\tilde{b}_{\ell_1 \ell_2 \ell_3} = \frac{b_{\ell_1 \ell_2 \ell_3}}{C_{\ell_1}C_{\ell_2} + C_{\ell_2}C_{\ell_3}+C_{\ell_3}C_{\ell_1}},
\ee
which reads in terms of the functions $R_{\ell}(r) $ and $\xi_{\ell}(r)$
\be
\tilde{b}_{\ell_1 \ell_2 \ell_3} = 8 \left( \frac{2}{\pi}\right)^{3/2} \frac{\nu_2}{C_{\ell_1}C_{\ell_2} + C_{\ell_2}C_{\ell_3}+C_{\ell_3}C_{\ell_1}} \,  \int {\rm d}r \left[ R_{\ell_1}(r) \xi_{\ell_2}(r) \xi_{\ell_3}(r) +  R_{\ell_2}(r) \xi_{\ell_3}(r) \xi_{\ell_1}(r) + R_{\ell_3}(r) \xi_{\ell_1}(r) \xi_{\ell_2}(r) \right] \; .
\ee

\subsubsection{Small angle approximation}

The reduced bispectrum encodes all the physical processes that lead to a non vanishing bispectrum. On the other hand, the Gaunt integral that appears in Eq.(\ref{fullbis}) only carries an overall geometrical dependence and ensures that the momenta $\ell_1, \ell_2$ and $\ell_3$ satisfy the triangular inequalities.

It appears that this overall geometrical dependence translates into a simple momentum conservation in the flat sky approximation (see Appendix \ref{sec:flatsky}). Defining the quantity $a(\vl)$ as in Appendix \ref{sec:flatsky}, the bispectrum reads in the small angle approximation
\be
\langle a(\vl_1) a(\vl_2) a(\vl_3) \rangle_{\rm c} = \frac{1}{2 \pi} b_{\ell_1 \ell_2 \ell_3} \delta_{\rm Dirac}(\vl_1+\vl_2+\vl_3)
\ee 
where $b_{\ell_1 \ell_2 \ell_3}$ is the reduced bispectrum defined in Eq.~(\ref{fullbis}). The Dirac function imposes that the vectors $\vl_1, \vl_2$ and $\vl_3$ form a triangle whose lengths respectively match with $\ell_1, \ell_2$ and $\ell_3$.

We now turn to the study of the tri-spectrum.

\begin{figure} 
\centerline{ \epsfig {figure=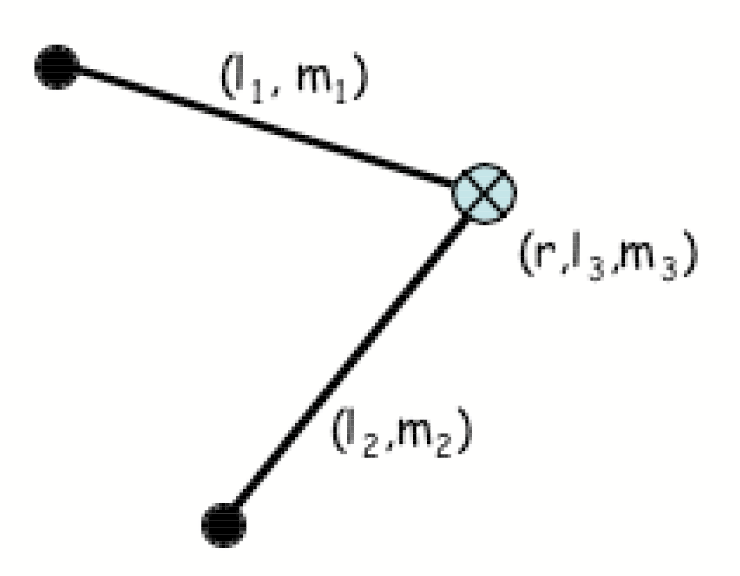,width=6cm,height=5cm}} 
\caption{Diagrammatic representation of the third term in the expression~(\ref{bisexpr}) of the bispectrum. Each line between a non-vertex point $(\ell_i,m_i)$ and a vertex $(r,\ell_3,m_3)$ represents the term $\xi_{\ell_i}(r)$. The vertex value is $\nu_2 {}^{(3)} {\cal G}^{m_1 \, m_2 \, m_3}_{\ell_1 \, \ell_2 \, \ell_3}\; R_{\ell_3}(r)$. The radial vertex position should be integrated over and the whole graph should be multiplied by $8 \left( \frac{ 2}{ \pi} \right)^{3/2}$.} 
\label{bispectre} 
\end{figure}

\subsection{Trispectrum} 

\subsubsection{Full sky expression}

Generally the trispectrum reads 
\bea 
\label{alm4} 
\langle a_{\ell_1 m_1} a_{\ell_2 m_2} a_{\ell_3 m_3}  a_{\ell_4 m_4}\rangle = \left( \frac{ 2}{ \pi} \right)^4 &&\int {\rm d}r_1 \int {\rm d}r_2 \int {\rm d}r_3 \int {\rm d}r_4 R_{\ell_1}(r_1) R_{\ell_2}(r_2) R_{\ell_3}(r_3)  R_{\ell_4}(r_4) \nonumber\\ 
&&\langle \Phi_{\ell_1 m_1}(r_1) \Phi_{\ell_2 m_2}(r_2) \Phi_{\ell_3 m_3}(r_3) \Phi_{\ell_4 m_4}(r_4) \rangle \;. 
\eea 

From the expression (\ref{Trispec}), we will have 2 contributions. One \emph{star} contribution, 
$\langle a_{\ell_1 m_1} \dots  a_{\ell_4 m_4}\rangle^{\rm star}$ due to the first term and one \emph{line} contribution, $\langle a_{\ell_1 m_1} \dots  a_{\ell_4 m_4}\rangle^{\rm line}$, due to the second one. These denominations will actually become much clearer in section~\ref{sec:diagram} where a diagrammatic representation of those terms is developed. For instance, only the diagrams whose shape is drawn in Fig.~\ref{trispectre_line} contribute to the {\em line} trispectrum, whereas the {\em star} trispectrum is made of terms whose shape is drawn in Fig.~\ref{trispectre_star}.

For the line contribution, the connected four-point function of the potential is given by
\bea 
\langle \Phi(\vk_1) \dots \Phi(\vk_4) \rangle_c^{\rm line} &=& \nul \int {\rm d}^3\vk_5 \; P(k_5) \left[ P(k_1)+P(k_2)\right]\left[ P(k_3)+ P(k_4)\right] \; \delta_{\rm Dirac} \left( \vk_1+\vk_2-\vk_5 \right) \delta_{\rm Dirac} \left( \vk_3+\vk_4+\vk_5 \right)\nonumber \\
&&+ \left(2\leftrightarrow 3\right) + \left(2\leftrightarrow 4\right)\;, 
\eea 
which entails using Eq.~(\ref{diracbes}), 
\bea 
\langle \Phi_{\ell_1 m_1}(r_1) \Phi_{\ell_2 m_2}(r_2) \Phi_{\ell_3 m_3}(r_3) \Phi_{\ell_4 m_4}(r_4) \rangle_c^{\rm line} &=& \nul \frac{2^8}{\pi^2} \sum_{L,M} (-1)^M {\cal G}_{\ell_1 \, \ell_2 \, L}^{m_1  m_2 M} {\cal G}_{\ell_3 \, \ell_4 \, L}^{m_3  m_4 -M} \int {\rm d}x_1 {\rm d}x_2  \, x_1^2 \, x_2^2 \nonumber\\ 
&&\Bigg\{ \int {\rm d}k_1 \ k_1^2 P(k_1) j_{\ell_1}(k_1 r_1) j_{\ell_1}(k_1 x_1) \int {\rm d}k_2 \, k_2^2 j_{\ell_2}(k_2 r_2)  j_{\ell_2}(k_2 x_1)  \nonumber\\ 
&&\int {\rm d}k_3  \, k_3^2  P(k_3) j_{\ell_3}(k_3 r_3)  j_{\ell_3}(k_3 x_2) \int {\rm d}k_4  \, k_4^2  j_{\ell_4}(k_4 r_4) j_{\ell_4}(k_4 x_2) \nonumber\\  
&&\int {\rm d}k_5  \, k_5^2 P(k_5)  j_L(k_5 x_1) \, j_L(k_5 x_2) +\left(1\leftrightarrow 2\right) + \left(3\leftrightarrow 4\right)  + \left(\begin{array}{c}1\leftrightarrow 2\\ 3\leftrightarrow 4 \end{array}\right) \Bigg\}\nonumber\\
&& + \left(2\leftrightarrow 3\right) + \left(2\leftrightarrow 4\right)\; . 
\eea 
Integrations over $k_2$ and $k_4$ in the first term impose $x_1=r_2$ and $x_2=r_4$ such as 
\bea 
\label{phi4tri} 
\langle \Phi_{\ell_1 m_1}(r_1) \Phi_{\ell_2 m_2}(r_2) \Phi_{\ell_3 m_3}(r_3) \Phi_{\ell_4 m_4}(r_4) \rangle_c^{\rm line} &=& \nul 2^6 \sum_{L,M} (-1)^M {\cal G}_{\ell_1 \, \ell_2 \, L}^{m_1  m_2 M} {\cal G}_{\ell_3 \, \ell_4 \, L}^{m_3  m_4 -M} \int {\rm d}k_1 \, k_1^2 P(k_1) j_{\ell_1}(k_1 r_1)  j_{\ell_1}(k_1 r_2) \nonumber\\ 
&&\int {\rm d}k_3  \, k_3^2 P(k_3)  j_{\ell_3}(k_3 r_3) j_{\ell_3}(k_3 r_4) \int {\rm d}k_5 \, k_5^2 P(k_5) j_L(k_5 r_2) j_L(k_5 r_4) \nonumber\\ 
&&+\mbox{ 11 other terms.} \; . 
\eea 

Using the definition of the radial transfer function (\ref{radexpress}), integrations over two of the radial variables may be performed within the expression of the trispectrum which becomes 
\bea 
\label{triexpr} 
\langle a_{\ell_1 m_1} a_{\ell_2 m_2} a_{\ell_3 m_3}  a_{\ell_4 m_4}\rangle_c^{\rm line} &=&  8^2 \nul  \left( \frac{ 2}{ \pi} \right)^2 \sum_{L,M} (-1)^M {\cal G}_{\ell_1 \, \ell_2 \, L}^{m_1  m_2 M} {\cal G}_{\ell_3 \, \ell_4 \, L}^{m_3  m_4 -M} \nonumber \\ 
&& \Bigg\{ \int {\rm d}r \int {\rm d}r' R_{\ell_1}(r) \xi_{\ell_2}(r) \zeta_{L}(r,r') R_{\ell_3}(r') \xi_{\ell_4}(r') +\left(1\leftrightarrow 2\right) + \left(3\leftrightarrow 4\right)  + \left(\begin{array}{c}1\leftrightarrow 2\\ 3\leftrightarrow 4 \end{array}\right)  \Bigg\}\nonumber\\
&& +\left(2\leftrightarrow 3\right) + \left(2\leftrightarrow 4\right), 
\eea 
where 
\be 
\zeta_{L}(r,r')=\int {\rm d}k k^2 P(k) j_L(k r) j_L(k r')\; . 
\ee 
Contrarily to the previous correlation functions, the expression of the trispectrum involves a third function which only depends on the primordial power spectrum. We will see in section~\ref{sec:diagram} that this term has a diagrammatic interpretation. For instance, the second term in Eq.~(\ref{triexpr}) is diagrammatically represented in Fig.~\ref{trispectre_line}. 
 
For the \emph{star} contribution we have,
\bea 
\langle \Phi_{\ell_1 m_1}(r_1)  \Phi_{\ell_2 m_2}(r_2) \Phi_{\ell_3 m_3}(r_3) \Phi_{\ell_4 m_4}(r_4)\rangle_c^{\rm star} &=& \frac{2^7}{\pi} \, \nus \, \int {\rm d}^2 \hg Y^*_{\ell_1 m_1}(\hg) Y^*_{\ell_2 m_2}(\hg) Y^*_{\ell_3 m_3}(\hg) Y^*_{\ell_4 m_4}(\hg) \int {\rm d}x \, x^2 \nonumber\\ 
&& \int {\rm d}k_1 \, k_1^2 j_{\ell_1}(k_1 r_1) j_{\ell_1}(k_1 x) \int {\rm d}k_2 \, k_2^2 j_{\ell_2}(k_2 r_2) j_{\ell_2}(k_2 x)\nonumber\\ 
&& \int {\rm d}k_3 \, k_3^2 j_{\ell_3}(k_3 r_3) j_{\ell_3}(k_3 x) \int {\rm d}k_4 \, k_4^2 j_{\ell_4}(k_4 r_4) j_{\ell_4}(k_4 x)\nonumber\\ 
&&\left[ P(k_1) P(k_2) P(k_3) + \mbox{3 other terms} \right]\, . 
\eea 
The first term should be computed by integrating first over $k_4$, which imposes $x=r_4$. This gives 
\bea 
\label{phi4qua} 
\langle \Phi_{\ell_1 m_1}(r_1)  \Phi_{\ell_2 m_2}(r_2) \Phi_{\ell_3 m_3}(r_3) \Phi_{\ell_4 m_4}(r_4)\rangle_c^{\rm star} &=& 2^6 \, \nus \, \int {\rm d}^2 \hg Y^*_{\ell_1 m_1}(\hg) Y^*_{\ell_2 m_2}(\hg) Y^*_{\ell_3 m_3}(\hg) Y^*_{\ell_4 m_4}(\hg) \nonumber\\ 
&& \int {\rm d}k_1 \, k_1^2 P(k_1) j_{\ell_1}(k_1 r_1) j_{\ell_1}(k_1 r_4) \int {\rm d}k_2 \, k_2^2  P(k_2) j_{\ell_2}(k_2 r_2) j_{\ell_2}(k_2 r_4)\nonumber\\ 
&& \int {\rm d}k_3 \, k_3^2 P(k_3) j_{\ell_3}(k_3 r_3) j_{\ell_3}(k_3 r_4) +\mbox{3 other terms}\, . 
\eea 
  
The angular integral may be expressed as a function of Gaunt integrals 
\be 
\label{Gaunt4} 
\Gaunt{4}_{\ell_1 \, \ell_2 \, \ell_3\, \ell_4}^{m_1  m_2 m_3 m_4}
=\int {\rm d}^2 \hg Y_{\ell_1 m_1}(\hg) Y_{\ell_2 m_2}(\hg) Y_{\ell_3 m_3}(\hg) Y_{\ell_4 m_4}(\hg) = \sum_{L,M} (-1)^M \,{}^{(3)}{\cal G}^{m_1 \, m_2 \, M}_{\ell_1 \, \ell_2 \, L} \,{}^{(3)} {\cal G}^{m_3 \, m_4 \, -M}_{\ell_3 \, \ell_4 \, L}\; . 
\ee 
 
Using the same expansion as in the previous subsection, we get 
\be 
\label{triexpr4}
\langle a_{\ell_1 m_1} a_{\ell_2 m_2} a_{\ell_3 m_3}  a_{\ell_4 m_4} \rangle_c^{\rm star} = \nus \frac{2^7}{ \pi} \sum_{L,M} \,{}^{(3)} {\cal G}_{\ell_1 \, \ell_2 \, L}^{m_1  m_2 M}  \,{}^{(3)}{\cal G}_{\ell_3 \, \ell_4 \, L}^{m_3  m_4 -M} (-1)^M \int {\rm d}r \left[ R_{\ell_1}(r) \xi_{\ell_2}(r) \xi_{\ell_3}(r) \xi_{\ell_4}(r) + \mbox{3 other terms}\right] .
\ee 
One of the terms in Eq.~(\ref{triexpr4}) is diagrammatically represented in Fig.~\ref{trispectre_star}.

The expressions~\ref{triexpr} and~\ref{triexpr4} are in perfect agreement with \cite{HuTrispectre2} in the case of the trispectrum. Indeed, this formalism can easily be extended to higher order terms. Any N-point correlation function
may be expressed in terms of the three functions $R_{\ell}$, $\xi_{\ell}$ and $\zeta_L$ using some basic rules that are detailed in section \ref{sec:diagram} and associated with a diagrammatic description.

Before going further, we stress how to decouple the overall geometrical dependence given in terms of Gaunt integals from the primordial signal. To do so, we build an estimator for the full-sky trispectrum and study its limit in the small angle approximation.

\subsubsection{Estimator}

Following \cite{HuTrispectre1,TheseKomatsu}, we write the trispectrum in a rotational invariant form
\be
\langle a_{\ell_1 m_1} a_{\ell_2 m_2} a_{\ell_3 m_3}  a_{\ell_4 m_4}\rangle_{c} =  \sum_{L,M} (-1)^M \left( \begin{array}{crcl} \ell_1 & \ell_2 & L \\ m_1 & m_2 & M  \end{array}\right) \left( \begin{array}{crcl} \ell_3 & \ell_4 & L \\ m_3 & m_4 & -M  \end{array}\right)\;{\cal T}_{\ell_3\, \ell_4}^{\ell_1\, \ell_2}(L)\; .
\ee
Similarly to the bispectrum, an estimator for the connected part of the angular averaged trispectrum may be chosen to be

\bea
\label{EstimTri}
T_{\ell_3 \ell_4}^{\ell_1 \ell_2}(L) = (2L+1) \sum_{{\rm all}\; m} \sum_{M} (-1)^M \left( \begin{array}{crcl} \ell_1 & \ell_2 & L \\ m_1 & m_2 & M  \end{array}\right) \left( \begin{array}{crcl} \ell_3 & \ell_4 & L \\ m_3 & m_4 & -M  \end{array}\right)  a_{\ell_1 m_1} a_{\ell_2 m_2} a_{\ell_3 m_3}  a_{\ell_4 m_4} - G_{\ell_3 \ell_4}^{\ell_1 \ell_2}(L),
\eea
where $G_{\ell_3 \ell_4}^{\ell_1 \ell_2}(L)$ is the estimator of the unconnected terms and is defined in such a way that $T_{\ell_3 \ell_4}^{\ell_1 \ell_2}(L)$ vanishes for a Gaussian field (see ~\cite{HuTrispectre1} for a discussion though the observable is not explicitly written in there),

\bea
G_{\ell_3 \ell_4}^{\ell_1 \ell_2}(L) &=&  (2L+1) \sum_{{\rm all}\; m} \sum_{M} (-1)^M \left( \begin{array}{crcl} \ell_1 & \ell_2 & L \\ m_1 & m_2 & M  \end{array}\right) \left( \begin{array}{crcl} \ell_3 & \ell_4 & L \\ m_3 & m_4 & -M  \end{array}\right) \nonumber\\
&& \Bigg\{ \frac{(-1)^{m_1+m_3}}{(2\ell_1+1)(2\ell_3+1+2\delta_{\ell_{1}\ell_{3}})} \delta_{m_1 -m_2} \delta_{m_3 -m_4} \delta_{\ell_1 \ell_2} \delta_{\ell_3 \ell_4} \sum_{m',m''}  (-1)^{m'+m''} a_{\ell_1 m'} a_{\ell_1 -m'} a_{\ell_3 m''}  a_{\ell_3 -m''}  \nonumber\\
&&  + \left( 2 \leftrightarrow 3 \right) + \left( 2 \leftrightarrow 4 \right) \Bigg\}.
\eea

In the star configuration, we define the reduced averaged trispectrum $^{\rm star}t_{\ell_3 \ell_4}^{\ell_1 \ell_2}(L)$ as
\be
^{\rm star}{\cal T}_{\ell_3 \ell_4}^{\ell_1 \ell_2}(L) = \frac{2L+1}{4\pi}\sqrt{(2\ell_1+1)(2\ell_2+1)(2\ell_3+1)(2\ell_4+1)} \left( \begin{array}{crcl} \ell_1 & \ell_2 & L \\ 0 & 0 & 0  \end{array}\right) \left( \begin{array}{crcl} \ell_3 & \ell_4 & L \\ 0 & 0 & 0  \end{array}\right) \; ^{\rm star}t_{\ell_3 \ell_4}^{\ell_1 \ell_2}(L)\; . 
\ee
so that it takes the $L$-independent form
\be
{}^{\rm star}t_{\ell_3 \ell_4}^{\ell_1 \ell_2} =\nus \frac{2^7}{\pi} \int {\rm d}r \left[ R_{\ell_1}(r) \xi_{\ell_2}(r) \xi_{\ell_3}(r) \xi_{\ell_4}(r) + \mbox{3 other terms}\right] .
\ee

The case of the line configuration is much involved since it implies three kinds of terms with different geometrical configurations encoded by the Gaunt integrals. The line trispectrum takes the form  
\be
 \langle a_{\ell_1 m_1} a_{\ell_2 m_2} a_{\ell_3 m_3}  a_{\ell_4 m_4}\rangle_{c}^{\rm line} =  \sum_{L,M} (-1)^M \left( \begin{array}{crcl} \ell_1 & \ell_2 & L \\ m_1 & m_2 & M  \end{array}\right) \left( \begin{array}{crcl} \ell_3 & \ell_4 & L \\ m_3 & m_4 & -M  \end{array}\right)\;P_{\ell_3\, \ell_4}^{\ell_1\, \ell_2}(L)\nonumber\\
+ \left( 2 \leftrightarrow 3 \right) + \left( 2 \leftrightarrow 4 \right)\; ,
\ee
with
\be
P_{\ell_3 \ell_4}^{\ell_1 \ell_2}(L) = \frac{2L+1}{4\pi}\sqrt{(2\ell_1+1)(2\ell_2+1)(2\ell_3+1)(2\ell_4+1)} \left( \begin{array}{crcl} \ell_1 & \ell_2 & L \\ 0 & 0 & 0  \end{array}\right) \left( \begin{array}{crcl} \ell_3 & \ell_4 & L \\ 0 & 0 & 0  \end{array}\right) \; ^{\rm line} t_{\ell_3 \ell_4}^{\ell_1 \ell_2}(L)\; . 
\ee
and the reduced trispectrum $^{\rm line} t_{\ell_3 \ell_4}^{\ell_1 \ell_2}(L)$ is given by
\be
\label{eq:tline}
^{\rm line} t_{\ell_3 \ell_4}^{\ell_1 \ell_2}(L) = 8^2 \nul  \left( \frac{ 2}{ \pi} \right)^2 \Bigg\{ \int {\rm d}r \int {\rm d}r' R_{\ell_1}(r) \xi_{\ell_2}(r) \zeta_{L}(r,r') R_{\ell_3}(r') \xi_{\ell_4}(r') +\left(1\leftrightarrow 2\right) + \left(3\leftrightarrow 4\right)  + \left(\begin{array}{c}1\leftrightarrow 2\\ 3\leftrightarrow 4 \end{array}\right) \Bigg\}. 
\ee

Using the link between the Wigner-6j and Wigner-3j symbols (see \cite{HuTrispectre1,HuTrispectre2, Wigner6j})
\be
\left\{\begin{array}{crcl} a & b & e \\ c & d & f \end{array}\right\} = \sum_{\alpha \beta \gamma} \sum_{\delta \epsilon \phi} (-1)^{e+f+\epsilon+\phi} \left( \begin{array}{crcl} a & b & e \\ \alpha & \beta & \epsilon  \end{array}\right) \left( \begin{array}{crcl} c & d & e \\ \gamma & \delta & -\epsilon  \end{array}\right) \left( \begin{array}{crcl} a & d & f \\ \alpha & \delta & -\phi  \end{array}\right)  \left( \begin{array}{crcl} c & b & f \\ \gamma & \beta & \phi  \end{array}\right)\; ,
\ee
the quantity $^{\rm line}{\cal T}_{\ell_3\, \ell_4}^{\ell_1\, \ell_2}(L)$ can be written as
\bea
\label{6jtri}
^{\rm line}{\cal T}_{\ell_3\, \ell_4}^{\ell_1\, \ell_2}(L)& = & P_{\ell_3\, \ell_4}^{\ell_1\, \ell_2}(L)+(2L+1)\sum_{L'}(-1)^{\ell_2+\ell_3}\left\{\begin{array}{crcl} \ell_1 & \ell_2 & L \\ \ell_4 & \ell_3 & L' \end{array}\right\}  P_{\ell_2\, \ell_4}^{\ell_1\, \ell_3}(L')\nonumber\\
&&+(2L+1)\sum_{L''}(-1)^{L+L''}\left\{\begin{array}{crcl} \ell_1 & \ell_2 & L \\ \ell_3 & \ell_4 & L'' \end{array}\right\}  P_{\ell_3\, \ell_2}^{\ell_1\, \ell_4}(L'') \; .
\eea
The Wigner-6j symbol $\left\{\begin{array}{crcl} a & b & e \\ c & d & f \end{array}\right\}$  represents a quadrilateral with sides $(a,b,c,d)$ whose diagonal form the triangles $(a,d,f), (b,c,f), (c,d,e), (a,b,e)$ and vanishes if any of the related triangular inequalities are not fulfilled (see \cite{Wigner6j}). 

Note that contrary to the {\em star} trispectrum, the {\em line} trispectrum is generically singular when $L=0$. This is possible if $\ell_{1}=\ell_{2}$ and $\ell_{3}=\ell_{4}$ (and symmetric cases) and is due to the expression (\ref{eq:tline}) which is then logarithmically divergent for a scale invariant power spectrum. This apparent divergence however does not appear in observable quantities such that in Eq. (\ref{EstimTri}). Indeed it can be easily checked that terms involving $P_{\ell_3 \ell_4}^{\ell_1 \ell_2}(L=0)$ and coming from the two terms of  (\ref{EstimTri}) exactly cancel each other. Super-Hubble effects remain then unobservable.


Finally, we find convenient to introduce the normalized trispectra $\tilde{ \cal T}_{\ell_3\, \ell_4}^{\ell_1\, \ell_2}(L)$ defined as
\be
\label{NormTri}
\tilde{\cal T}_{\ell_3\, \ell_4}^{\ell_1\, \ell_2} = \frac{{\cal T}_{\ell_3\, \ell_4}^{\ell_1\, \ell_2}(L)}{(C_{\ell_1} +C_{\ell_2}) C_L (C_{\ell_3} +C_{\ell_4})}.
\ee

\subsubsection{Small angle approximation}

Using the notations of Appendix~\ref{sec:flatsky}, we can define the quantities $a(\vl)$ for large enough multipoles so that the line and star trispectra take the forms
\be
\langle a(\vl_1) a(\vl_2) a(\vl_3) a(\vl_4) \rangle_{\rm c}^{\rm star} = \frac{1}{(2\pi)^2} \delta_{\rm Dirac}(\vl_1+\vl_2+\vl_3+\vl_4) \; {}^{\rm star}t_{\ell_3\, \ell_4}^{\ell_1\, \ell_2} 
\ee
and
\bea
\langle a(\vl_1) a(\vl_2) a(\vl_3) a(\vl_4) \rangle_{\rm c}^{\rm line} &=& \frac{1}{(2\pi)^2} \delta_{\rm Dirac}(\vl_1+\vl_2+\vl_3+\vl_4) \int \dd^2 \vec{L} \; \left[ \delta_{\rm Dirac}(\vl_1+\vl_2-\vec{L}){}^{\rm line}t_{\ell_3\, \ell_4}^{\ell_1\, \ell_2}(L)\right. \nonumber\\
&& \left. \hspace{3cm} + \delta_{\rm Dirac}(\vl_1+\vl_3-\vec{L}){}^{\rm line}t_{\ell_2\, \ell_4}^{\ell_1\, \ell_3}(L)+  \delta_{\rm Dirac}(\vl_1+\vl_4-\vec{L}){}^{\rm line}t_{\ell_3\, \ell_2}^{\ell_1\, \ell_4}(L)\right] \; .
\eea
The Dirac functions ensures that the multipoles $\vl_1, \vl_2, \vl_3, \vl_4$ form a quadrilateral. The star-trispectrum does not depend on the shape of the quadrangle since it is enterely determined by the side lengths. Yet the line-trispectrum not only depends on the side lengths but also on the diagonals $\vec{L}_{12}=\vl_1+\vl_2$, $\vec{L}_{13}=\vl_1+\vl_3$ and $\vec{L}_{14}=\vl_1+\vl_4$ of the quadrilateral formed by the four multipoles.
  
\begin{figure} 
\centerline{ \epsfig {figure=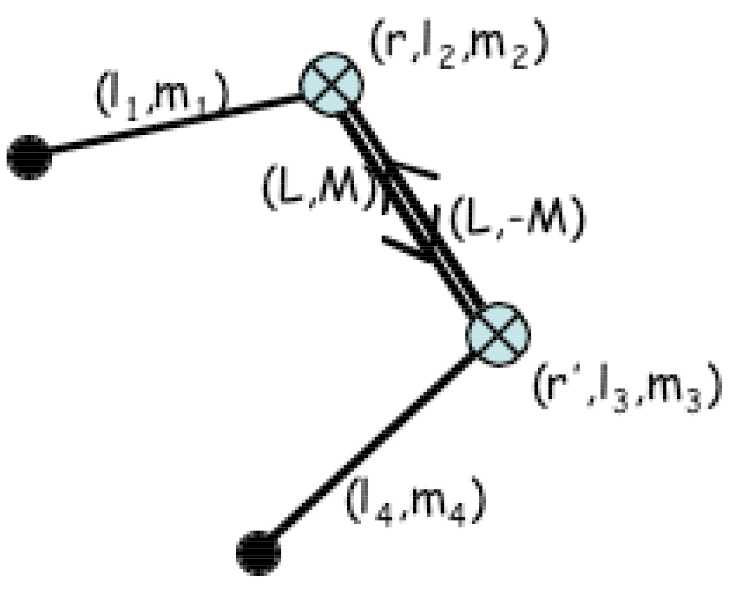,width=6cm,height=5cm}} 
\caption{Diagrammatic representation of the "line" contribution to the trispectrum. Each line between a non-vertex point $(\ell_i,m_i)$ and a vertex $(r,\ell,m)$ represents the term $\xi_{\ell_i}(r)$. Each line between the vertices $(r,\ell_2,m_2)$ and $(r',\ell_3,m_3)$ represents the term $\zeta_L(r,r')$. 
 (Here $L$ is the total angular momentum from $\ell_1$ and $\ell_2$ or equivalently from $\ell_3$ and $\ell_4$). The vertex values are $\nu_2 {}^{(3)} {\cal G}^{m_1 \, m_2 \, M}_{\ell_1 \, \ell_2 \, L}  \; R_{\ell_2}(r)$  and $\nu_2 {}^{(3)} {\cal G}^{m_3 \, m_4 \, -M}_{\ell_3 \, \ell_4 \, L} \; R_{\ell_3}(r')$. The radial positions of the vertices should be integrated over and the internal momenta $(L,M)$ should be summed over. The whole graph should be multiplied by $8^2 \left(\frac{2}{\pi}\right)^2$.} 
\label{trispectre_line} 
\end{figure}

\begin{figure} 
\centerline{ \epsfig {figure=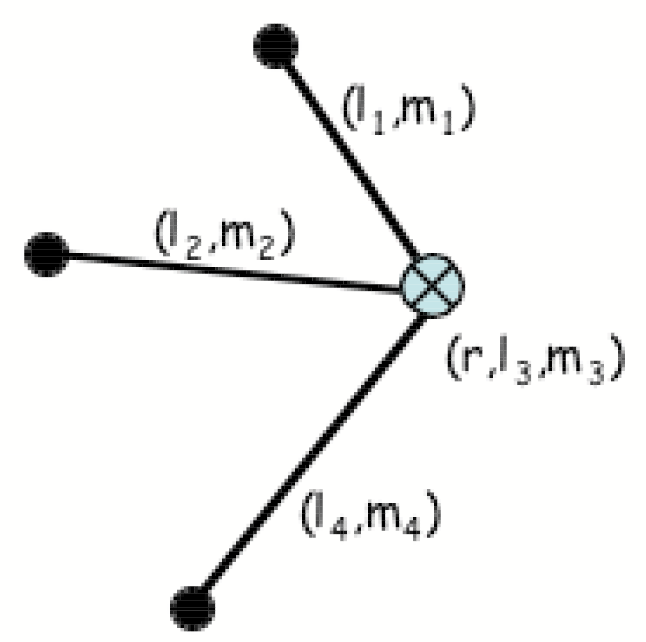,width=5cm,height=5cm}} 
\caption{Diagrammatic representation of the "star" contribution to the trispectrum. Each single line between $(\ell_i,m_i)$ and the vertex $(r,\ell,m)$ represents the term $\xi_{\ell_i}(r)$. The vertex value at the point $r$ is given by $\nu_3 {}^{(4)} {\cal G}^{m_1 \, m_2 \, m_3 \, m_4}_{\ell_1 \, \ell_2 \, \ell_3 \, \ell_4} \; R_{\ell_3}(r)$. The radial position of the vertex should be integrated over. The whole graph should be multiplied by $ \frac{ 2^7}{ \pi}$.} 
\label{trispectre_star} 
\end{figure}

\subsection{Diagrammatic representations}
\label{sec:diagram}

The formal results obtained in the previous sections show that at least some aspects of the potential correlation functions are preserved. That furthermore suggests general rules for the constructions of the N-point correlation functions for the spherical harmonics coefficients.
Let us start with basics rules for the potential correlation function. Using a usual diagrammatic representation, assume that the following rules hold for any potential $N$-point function $\langle \Phi_{\vec{k}_1}...\Phi_{\vec{k}_N}\rangle_c$ for tree order or loop graphs:
\begin{itemize}
\item each vertex connecting $n_i$ lines contributes to a factor $\nu_{n_i-1}$ and a Dirac distribution of the sum over all the connected momenta,
\item each external line carrying the momentum $k_i$ represents $P(k_i)$. For $p_i$ external lines connected to a vertex, the contribution is a sum over configurations of the product of $p_i-1$ potential power spectra $P(k_i)$,   
\item each link between two vertices $i$ and $j$ should be formed by internal line. Each internal line stands for $P(q_i)$ and imposes a momentum conservation through $\delta_{\rm Dirac}(q_i+q_j)$, where $q_i$ (respectively $q_j$) is the internal momentum out of the vertex $i$ (respectively $j$).  
\end{itemize}
To compute the induced $N$-point temperature correlation function, we use the basic relation
\bea
\label{almNpoint}
\langle a_{\ell_1 m_1}...a_{\ell_N m_N}\rangle &=& \left(\frac{2}{\pi}\right)^{3N/2} (-i)^{\ell_1+...+\ell_N} \int \dd r_1...\dd r_N \; R_{\ell_1}(r_1)... R_{\ell_N}(r_N)\nonumber\\
&& \int \dd^3 k_1...\dd^3 k_N \; \langle \Phi_{\vec{k}_1}...\Phi_{\vec{k}_N}\rangle\nonumber\\
&&j_{\ell_1}(k_1 r_1)...j_{\ell_N}(k_N r_N) \; Y^*_{\ell_1 m_1}(\hk_1)...Y^*_{\ell_N m_N}(\hk_N)\; .
\eea

In the expression of the potential $N$-point function, the momentum conservation at the vertex $i$ with $n_i-p_i$ external lines and $p_i$ internal lines contributes as
\bea
\label{NDirac}
\sum_{L^i_1,...,L^i_{n_i},M^i_1,...,M^i_{n_i}} i^{L^i_1+...+L^i_{n_i}} \frac{(4\pi)^{n_i}}{(2\pi)^3} &&\int \dd^3 x_i \; j_{L^i_1}(k_1 x_i)...j_{L^i_{n_i-p_i}}(k_{n_i-p_i} x_i)j_{L^i_{n_i-p_i+1}}(q_1 x_i)...j_{L^i_{n_i}}(q_{p_i} x_i)\nonumber\\
&&Y_{L^i_1M^i_1}(\hk_1)...Y_{L^i_{n_i-p_i}M^i_{n_i-p_i}}(\hk_{n_i-p_i})Y_{L^i_{n_i-p_i+1}M^i_{n_i-p_i+1}}(\hat{q}_1)...Y_{L^i_{n_i}M^i_{n_i}}(\hat{q}_{p_i})\nonumber\\
&&Y^*_{L^i_1M^i_1}(\hg_i)...Y^*_{L^i_{n_i}M^i_{n_i}}(\hg_i)\;, 
\eea
where $k$ ($q$) stands for an external (internal) momentum.

 Integration over the angles $\hg_i$ will contribute with the overall geometrical factor $\Gaunt{n_i}_{L^i_1... L^i_{n_i}}^{M^i_{1} M^i_{n_i}}$. Integration over the angles $\hk_i$ in Eq.~(\ref{almNpoint}) will impose $L^i_j=\ell_j$ for all external momenta. Integrations over all the external momenta that do not appear as $P(k_i)$ in the expansion Eq.~(\ref{almNpoint}) give $\pi/2 \; \delta_{\rm Dirac}(r_j-x_k)$ factors (see Eq.~(\ref{dirac})) which are to be integrated over $x_k$. Only the term
\be
\int \dd r_j R_{\ell_j}(r_j)
\ee 
remains. Those terms are induced by a vertex in the potential correlation function and may be associated with vertices at the positions $r_j$ that carry the momenta $(\ell_j,m_j)$. For all the other external momenta, integrations lead to
\be
\int \dd r_i \int \dd k_i \; k_i^2 P(k_i) R_{\ell_i}(r_i) j_{\ell_i}(k_ir_i) j_{\ell_i}(k_ir_k) \propto \xi_{\ell_i}(r_k)\; .
\ee
Those terms are easily understood as external lines attached to a vertex at the point $r_k$ as they correspond to external lines in the potential correlation function. Each external line carries the momenta $(\ell_i,m_i)$. One should note that $n_i$ lines attached to a vertex in the potential correlation function correspond to $n_i-1$ lines in the temperature correlation function plus a vertex which carries an angular momentum.

Integration over any internal momentum $q_i$ will give through $P(q_i) \delta_{\rm Dirac}(q_i+q_j)$ and Eq.~(\ref{NDirac}):
\be
\sum_{L,M}\zeta_L(x_i,x_j) Y^*_{L M}(\hg_i) Y_{L M}(\hg_j)\;, 
\ee
with $L=L^i_k=L^j_{k'}$ and $M=M^i_k=-M^j_{k'}$. These terms are involved as soon as an internal line connects two vertices in the potential correlation function. In a diagrammatic representation of the temperature correlation function, they may be represented by a double line (one carrying the azimuthal momentum $+M$, the other carrying $-M$) connecting the vertices located at $x_i$ and $x_j$.

Finally, if $N$ is the number of external points and $r$ is the number of vertex connecting $n_1,...,n_r$ lines respectively in the potential correlation function, the numerical prefactor $A_\alpha$ for the temperature correlation function reads
\be
A_\alpha =\nu_{n_1-1}...\nu_{n_r-1}\;2^{r+5N/2-4}\pi^{N/2-r-2}\; .
\ee
To summarize, the general temperature $N$-point function may be diagrammatically represented with the following rules deduced from the basic diagrammatic rules
\begin{itemize}
\item each of the $N$ harmonic coefficients $a_{\ell_i,m_i}$ is represented either by an external line or by a vertex with a charge $(\ell_i,m_i)$,
\item vertices are connected together by internal lines with arbitrary indices attached to them, being $(L_i,M_i)$ at one end and $(L_i,-M_i)$ at the other one,
\item one should then consider all the possible diagram configurations,
\end{itemize}
and the following correspondences allow the computation of any diagram
\begin{itemize}  
\item each external line with indices $(\ell_i,m_i)$ and attached to a vertex point at $r_j$ contibutes as $\xi_{\ell_i}(r_i)$,
\item each internal line with indices $(L_i,M_i)$ and $(L_i,-M_i)$ connecting two vertices at the points $r_i$ and $r_j$ stands for $\zeta_{L_i}(r_i,r_j)$,
\item the weight of a vertex with indices $(\ell_i,m_i)$ at the point $r_i$ to which $n_i$ lines of respective indices $(\ell^i_1,m^i_1),\dots,(\ell^i_{n_i},m^i_{n_i})$ are attached is $\nu_{n_i} \; R_{\ell_i}(r_i) \; \Gaunt{n_i+1} _{\ell^i_1 \, \dots \, \ell^i_{n_i} \,\ell_i}^{m^i_1 \dots m^i_{n_i} m_i}$,
\item if $N$ is the number of external points and $r$ is the number of vertices the numerical prefactor reads $\left(\frac{2}{\pi}\right)^{N/2} \frac{(4\pi)^{N+2r-2}}{(2\pi)^{3r}}=2^{r+5N/2-4}\pi^{N/2-r-2}$,
\item the final value of the diagram is obtained after integration over the radial variables and summation over in the internal indices $L_i$ and $M_i$.
\end{itemize}

Obviously, if those rules are applied to the diagrams of Figs.~\ref{bispectre},~ \ref{trispectre_line} and~\ref{trispectre_star}, one recovers the expressions~(\ref{bisexpr}), (\ref{triexpr}) and (\ref{triexpr4}). 

Those rules appear to be useful in the computation of any N-point correlation function in the temperature fluctuations. They provide
compact and simple expressions associated with diagrammatic interpretations. 

Having established the forms of the temperature correlation functions, we pay attention in the following section to the behaviors of the bi- and tri-spectra as functions of the multipoles and the configurations.

\section{Shapes of the correlation functions}

The features of high order correlation functions in the temperature fluctuations are difficult
to infer. However, at large angular scales, the temperature and potential fluctuations are related in a very simple way because low multipoles correspond to very few Fourier modes. Hence, in the Sachs-Wolfe limit (valid up to $\ell \sim 20$) the functional forms of the temperature and potential correlation functions should be the same. On the contrary, temperature fluctuations at small angular scales are induced by numerous Fourier modes of the potential: the initial distribution of potential fluctuations is altered by projection effects. This renders the functional form of the temperature correlation functions much more intricate. Before we explore general configurations, we first establish  the explicit Sachs-Wolfe limits of the bi- and tri-spectra and examine specific configurations. 

\subsection{The Sachs-Wolfe limits}

The Sachs-Wolfe limits of the functions $R_\ell$, $\xi_\ell$ read
\begin{eqnarray}
\label{RSW}
R_{\ell}^{\rm SW}(r)&=&-\frac{\pi}{6} \; \delta_{\rm Dirac}(r-\rs) \;, \\
\label{XSW}
\xi^{\rm SW}_{\ell}(\rs) &=& -\frac{3\pi}{2} \; C_{\ell}^{\rm SW} \;, 
\end{eqnarray}
whereas the vertex propagator contributes as
\be
\label{ZSW}
\zeta_L^{\rm SW}(\rs,\rs) = \int \dd k \; k^2P(k)j_L^2(k\rs)=\frac{9\pi}{2} \; C_L^{\rm SW}.
\ee

From Eqs.~(\ref{RSW})-(\ref{ZSW}), the Sachs-Wolfe limit of the reduced bispectrum takes the form
\be
b^{\rm SW}_{\ell_1 \ell_2 \ell_3} = -24 \nu_2 \left(\frac{\pi}{2}\right)^{3/2} \; \left( C_{\ell_1}^{\rm SW} C_{\ell_2}^{\rm SW} + {\rm perm.} \right),
\ee
with $\ell_1, \ell_2, \ell_3$ satisfying the triangular inequalities. As expected, the functional form of the temperature bispectrum is the same as the potential bispectrum. 

As the bispectrum scales as $\sim C_{\ell}^2$ at large angular scales, the normalized bispectrum (\ref{Normbis}) reduces to a constant in the Sachs-Wolfe limit
\be
\tilde{b}_{\ell_1 \ell_2 \ell_3}^{\rm SW} = -24 \nu_2 \left(\frac{\pi}{2}\right)^{3/2}\; .
\ee

Similar calculations can be done for the trispectra. In the line configuration, the reduced trispectrum takes the Sachs-Wolfe limit
\be
\label{TriSWline}
^{\rm line}t_{\ell_3 \ell_4}^{\ell_1 \ell_2}(L)^{\rm SW}  = 9 \nul ( 2\pi)^3 \; C_L^{\rm SW} \; \left( C_{\ell_1}^{\rm SW}+ C_{\ell_2}^{\rm SW}\right) \left(  C_{\ell_3}^{\rm SW}+ C_{\ell_4}^{\rm SW} \right) \;, 
\ee
while the $L$-independent star contribution gives
\be
\label{TriSWstar}
^{\rm star}t_{\ell_3 \ell_4}^{\ell_1 \ell_2}{}^{\rm SW} = 9 \nus \left(2\pi \right)^3 \left( C_{\ell_1}^{\rm SW} C_{\ell_2}^{\rm SW} C_{\ell_3}^{\rm SW}+ \mbox{3 other terms}\right) \; .
\ee
From Eqs.~(\ref{TriSWline}) and~(\ref{TriSWstar}), we recover the functional form of the potential tri-spectra when $\ell_1$, $\ell_2$ and $L$ on one hand and $\ell_3$, $\ell_4$ and $L$ on the other hand satisfy the triangle inequalities. One can note that the scaling of the temperature trispectrum at large angular scales depends on the configuration: in the line configuration, the trispectrum scales as $\sim C_l^2$ for a given $L$ whereas it scales as $\sim C_l^3$ in the star configuration. The shape of the temperature tri-spectrum then depends on the type of mode couplings that are responsible for the potential tri-spectrum. 
The normalized trispectra defined in Eq.~(\ref{NormTri}) reduce in the Sachs-Wolfe limit to 
\bea
^{\rm line} \tilde{t}_{\ell_3 \ell_4}^{\ell_1 \ell_2}(L) &=& 9 \nul \left(2 \pi\right)^3 \left[1+\frac{ \left( C_{\ell_1}+ C_{\ell_3} \right) C_{L_{13}} \left( C_{\ell_2} +C_{\ell_4} \right) }{ \left( C_{\ell_1}+ C_{\ell_2} \right) C_L \left( C_{\ell_3} +C_{\ell_4} \right) } + \frac{ \left( C_{\ell_1}+ C_{\ell_4} \right) C_{L_{14}} \left( C_{\ell_3} +C_{\ell_2} \right) }{ \left( C_{\ell_1}+ C_{\ell_2} \right) C_L \left( C_{\ell_3} +C_{\ell_4} \right)} \right] \;, \\
^{\rm star} \tilde{t}_{\ell_3 \ell_4}^{\ell_1 \ell_2} &=& 9 \nus \left( 2 \pi \right)^3  \frac{ C_{\ell_1} C_{\ell_2} C_{\ell_3} + \mbox{3 other terms}}{\left(C_{\ell_1}+ C_{\ell_2}\right) C_L \left(C_{\ell_3} +C_{\ell_4}\right)}\; . 
\eea

\subsection{General case}

In the previous paragraph it has been shown that the structure of the high order correlation functions for the temperature field reproduces those of the potential in the Sachs Wolfe limit. In general however this is not the case and the temperature correlation functions exhibit acoustic oscillations that depend on the geometric configuration~\cite{KS}. 

\begin{figure}
\centerline{
\begin{tabular}{cc} 
\epsfig {figure=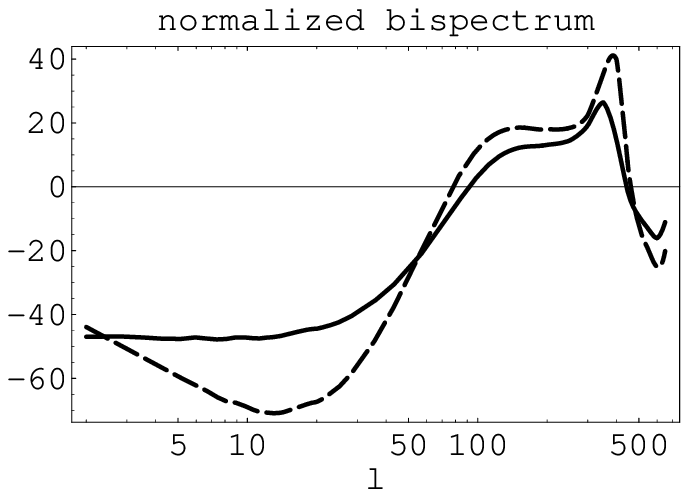, width=7cm,height=4.5cm} & \epsfig {figure=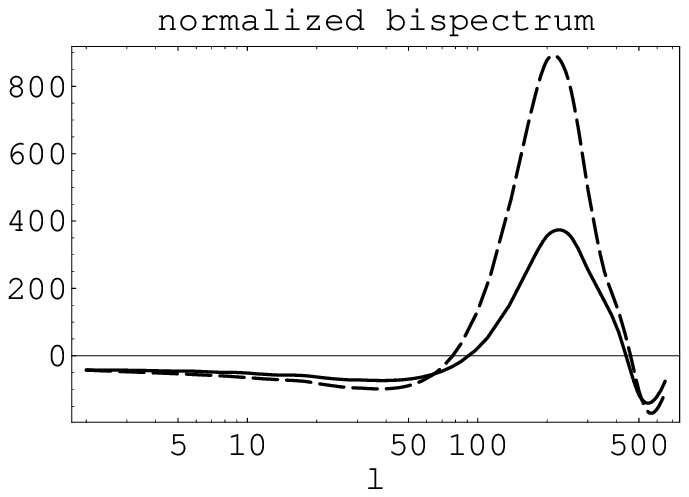, width=7cm,height=4.5cm}  
\end{tabular} }
\caption{Left panel: normalized bispectrum for an equilateral configuration $\tilde{b}_{\ell \ell \ell}$ in a $\Lambda$CDM model (dashed line) and a sCDM model (solid line) for $\nu_2=1$. The integrated Sachs Wolfe effect prevents the formation of a plateau at low $\ell$ in a $\Lambda$CDM model. Both curves exhibit acoustic oscillations of roughly the same amplitude although acoutic pics seem to be larger in a $\Lambda$CDM model. Right panel: $\ell^2 (\ell+1)^2 b_{\ell \ell \ell}$ in a $\Lambda$CDM model (dashed line) and a sCDM model (solid line) for $\nu_2=1$. The amplitude of the bispectrum is greater in a $\Lambda$CDM model. Comparison between the two plots shows that the second peak ($\ell\sim 300$) in $\tilde{b}_{\ell \ell \ell}$ is only due to the minimum of the $C_\ell$'s.} 
\label{fig:equi} 
\end{figure}

For example, left panel of Fig~\ref{fig:equi} shows the normalized bispectrum for an equilateral configuration in a $\Lambda$-cold dark matter model ($\Lambda$CDM) and a standard cold dark matter model (sCDM). A Sachs Wolfe plateau is expected in a sCDM model while in a $\Lambda$CDM model, the integrated Sachs-Wolfe effect strongly contributes to the bispectrum at low $\ell$ in such a way that the plateau is not visible any more. 
For greater $\ell$, say $\ell>20$, we can see the imprints of acoustic oscillations on the behaviors of the bispectra which vanish for some values of $\ell$. The amplitudes of oscillations of the normalized bispectra are comparable. As for the $C_\ell$'s, the period of acoutic oscillations seems to be larger in a $\Lambda$CDM model.
The right panel of Fig~\ref{fig:equi} shows the bispectrum with a different normalization, i.e. $\ell^2(\ell+1)^2 b_{\ell \ell \ell}$. We can note the presence of a plateau at the same level for both bispectra in the low $\ell$ limit although the amplitude of oscillations is greater in a $\Lambda$CDM model. Comparing the two normalizations, we can interpret the secondary peak of the normalized bispectrum $\tilde{b}_{\ell \ell \ell}$ at $\ell\sim 300$ as due to the minimum of the $C_\ell$'s.

\begin{figure}
\centerline{
\begin{tabular}{ccc}
\epsfig {figure=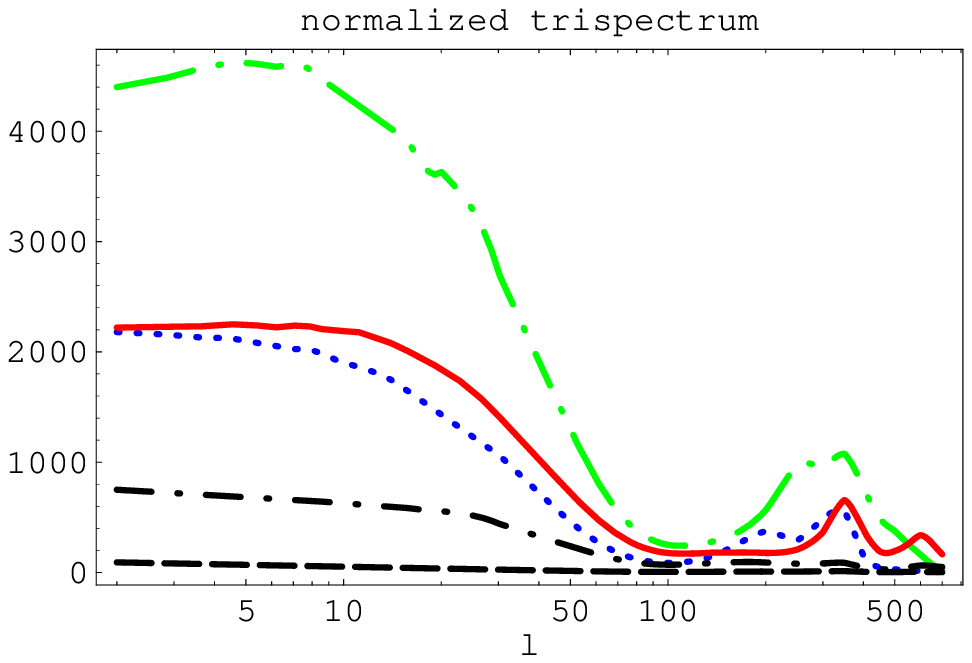, width=6cm,height=4cm } & \epsfig {figure=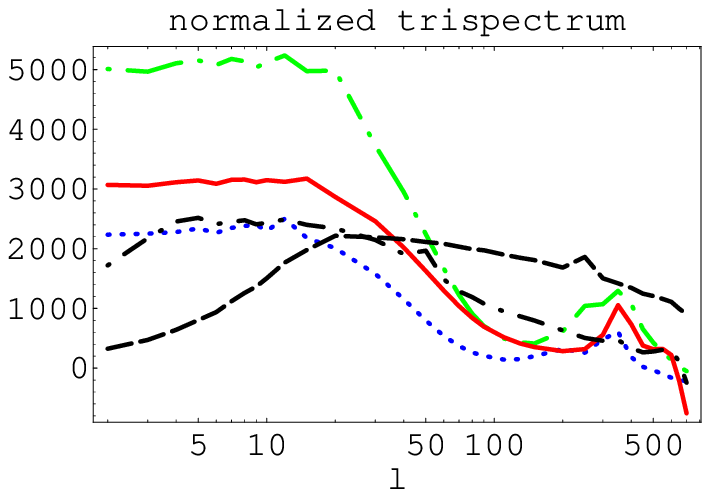, width=6cm,height=4cm } & \epsfig {figure=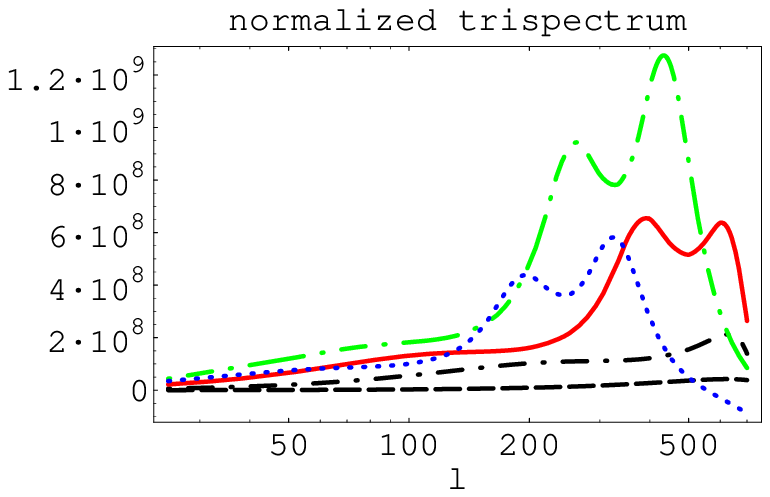, width=6cm,height=4cm } 
\end{tabular} }
\caption{Plot of ${\tilde t}^{\ell  \ell}_{\ell \ell}(L)$ as a function of $\ell$ for $L=\ell/10$ (dashed line), $L=\ell/2$ (dash-dotted line), $L=\ell$ (solid red line), $L=3 \ell/2$ (long-dash-dotted green line) and $L=2 \ell$ (dotted blue line) for $\nul = \nus =1$. The left panel represents the {\em star} trispectrum, the center panel shows the {\em line} trispectrum and the trispectrum from lensing effects is plotted in the right panel.}
\label{fig:TriEqui}
\end{figure}

\begin{figure}
\centerline{
\begin{tabular}{ccc}
\epsfig {figure=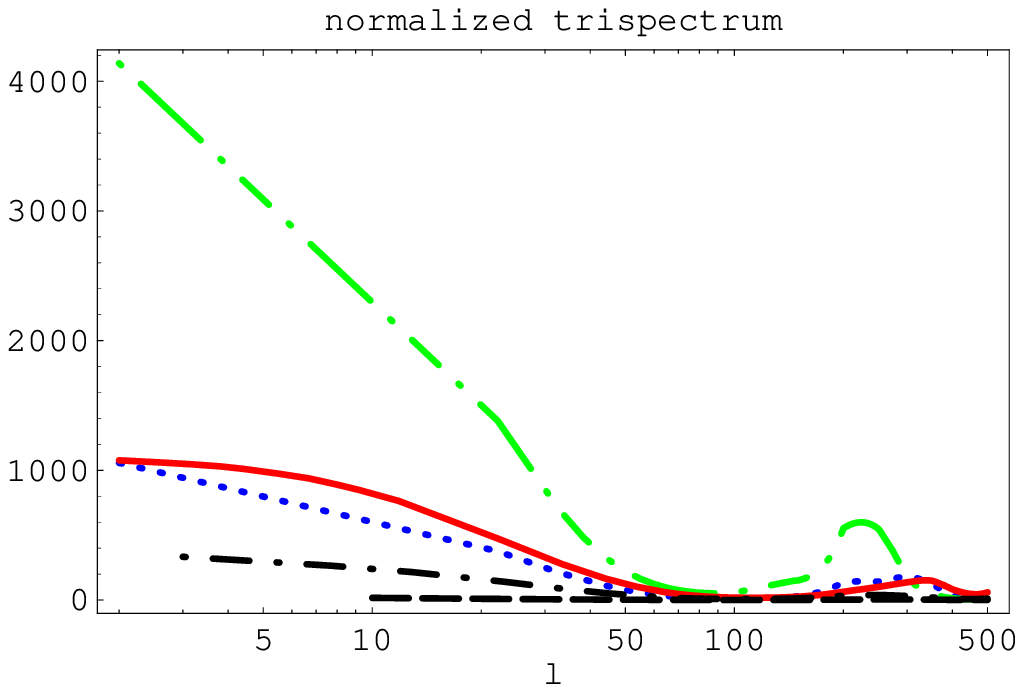, width=6cm,height=4cm } & \epsfig {figure=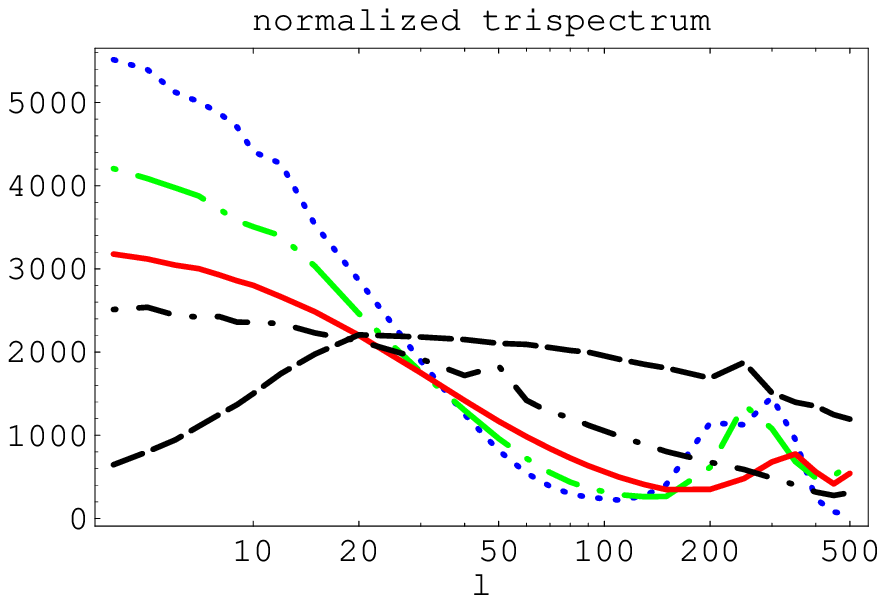, width=6cm,height=4cm } & \epsfig {figure=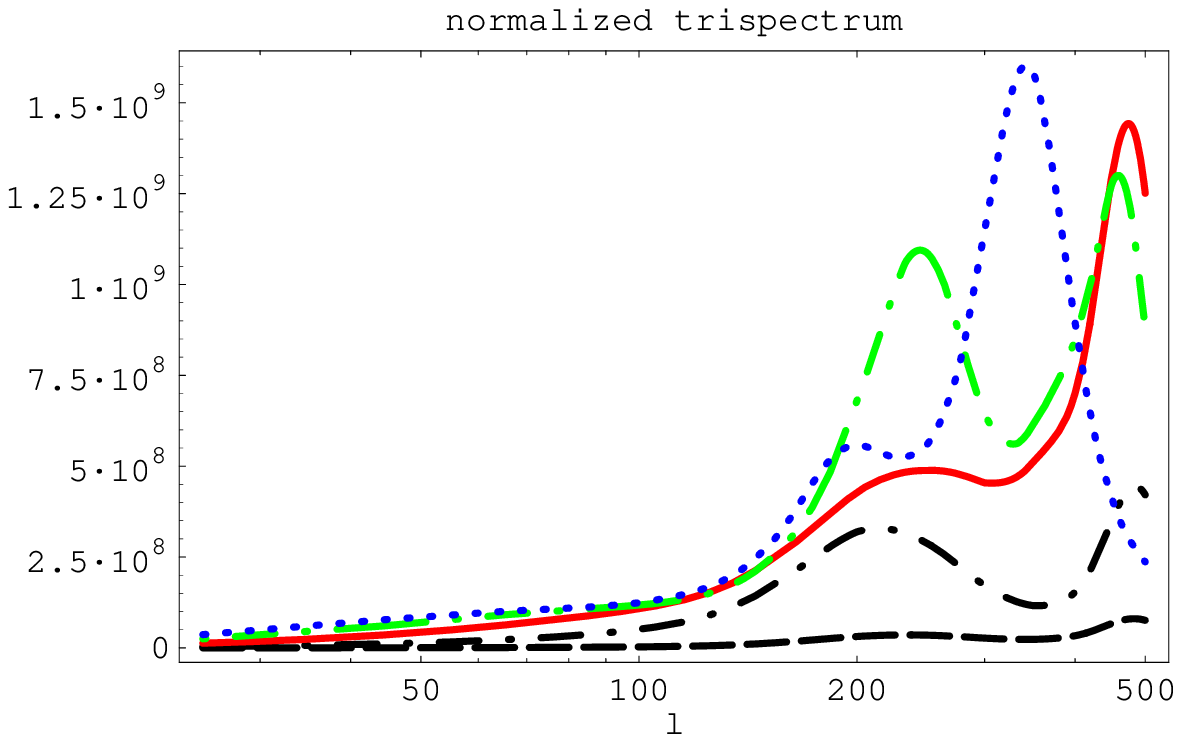, width=6cm,height=4cm } 
\end{tabular} }
\caption{Plot of ${\tilde t}^{\ell \;  \ell}_{\ell+200 \, \ell+200}(L)$ as a function of $\ell$ for $L=\ell/10$ (dashed line), $L=\ell/2$ (dash-dotted line), $L=\ell$ (solid red line), $L=3 \ell/2$ (long-dash-dotted green line) and $L=2 \ell$ (dotted blue line) for $\nul = \nus =1$. The left panel represents the {\em star} trispectrum, the center panel shows the {\em line} trispectrum and the trispectrum from lensing effects is plotted in the right panel.}
\label{fig:Tri200}
\end{figure}

\begin{figure}
\centerline{
\begin{tabular}{ccc}
\epsfig {figure=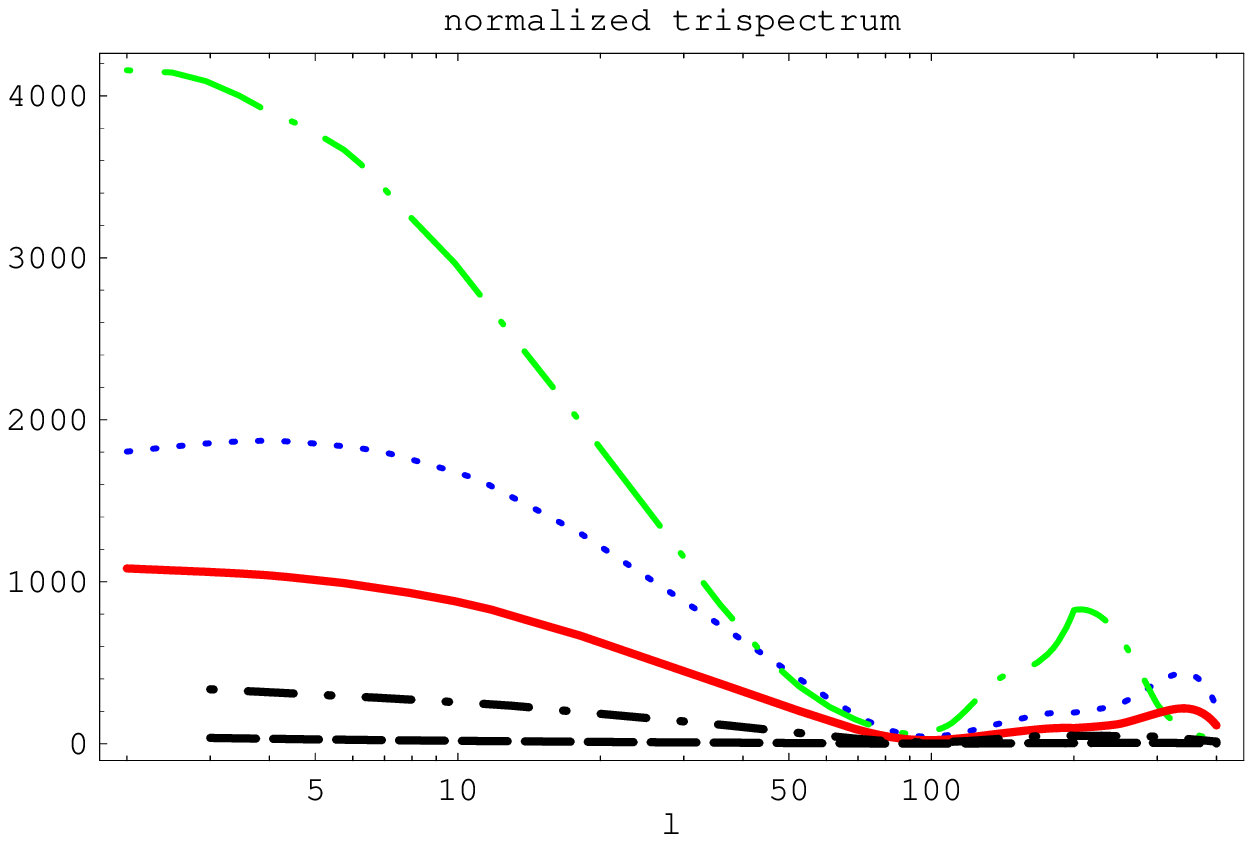, width=6cm,height=4cm } & \epsfig {figure=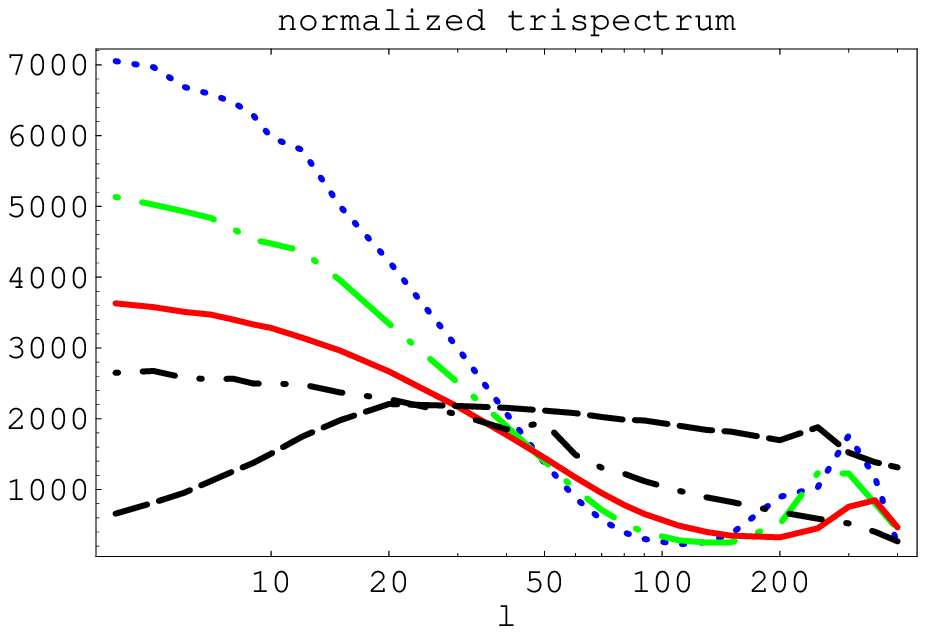, width=6cm,height=4cm } & \epsfig {figure=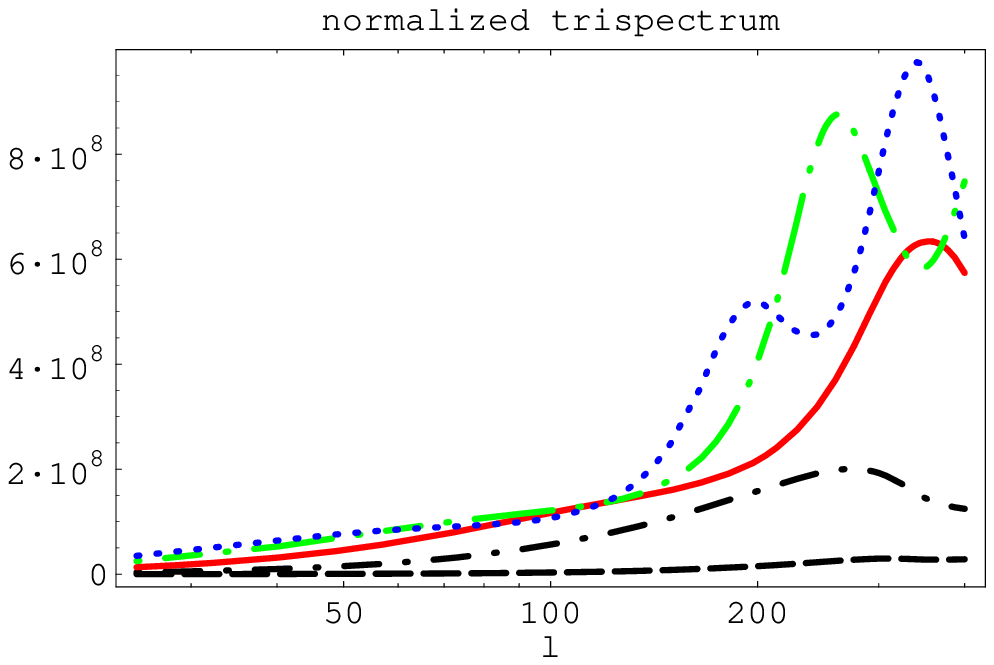, width=6cm,height=4cm } 
\end{tabular} }
\caption{Plot of ${\tilde t}^{\ell \;  \ell}_{\ell+300 \, \ell+300}(L)$ as a function of $\ell$ for $L=\ell/10$ (dashed line), $L=\ell/2$ (dash-dotted line), $L=\ell$ (solid red line), $L=3 \ell/2$ (long-dash-dotted green line) and $L=2 \ell$ (dotted blue line) for $\nul = \nus =1$. The left panel represents the {\em star} trispectrum, the center panel shows the {\em line} trispectrum and the trispectrum from lensing effects is plotted in the right panel.}
\label{fig:Tri300}
\end{figure}

We have similarly computed $\ttl$ and $\tts$ for different configurations. Some of those results are displayed in Figs.~\ref{fig:TriEqui},~\ref{fig:Tri200} and~\ref{fig:Tri300} for the {\em star} (left panels) and {\em line} like mode couplings (center panels).
  
Fig.~\ref{fig:TriEqui} corresponds to the losange configuration $\ell_1=\ell_2=\ell_3=\ell_4=\ell$ whereas Figs.~\ref{fig:Tri200} and~\ref{fig:Tri300} respectively correspond to the configurations $\ell_1=\ell_2=\ell$, $\ell_3=\ell_4=\ell+200$ and $\ell_1=\ell_2=\ell$, $\ell_3=\ell_4=\ell+300$. The different curves correspond to different losange shapes: $L=\ell/10$, $L=\ell/2$, $L=\ell$, $L=3\ell/2$ and $L=2 \ell$. 
We can see very similar behaviors of the trispectra: they all exhibit oscillations with an acoustic peak located at $\ell\sim 300$. The {\em line} and {\em star} signals are larger in the low $\ell$ range and have roughly the same order of magnitude when $\nul \sim \nus$. In the chosen configurations, the trispectra both take positive values at least for $\ell<500$. 

However, the {\em line} trispectrum is much larger than the {\em star} one for large $\ell$ as well as for low-$L$ values. A significant and specific dependence on the shape of the quadrilateral is clearly visible in these plots and may allow to distinguish between the two types of mode couplings. In Figs.~\ref{fig:Tri200} and~\ref{fig:Tri300}, we can see the $L$-dependence of the line trispectrum: for $\ell<30$, the signal is an increasing function of $L$ while it decreases with $L$ in the range $30<\ell<150$. This feature is quite different from the star trispectrum whose signal is not a monotonic function of $L$.  

The chosen configurations exhibit special points: for $\ell\sim 30$ for the line like mode coupling and $\ell \sim 80$ for the star like one, the trispectra take values that do not depend on $L$. Note that for these points, the value of the star trispectrum is very low contrary to the line trispectrum. However, we should stress that we only plotted some specific configurations whose features may be quite different from other geometrical configurations. Exploring all the configurations remain a daunting task although we expect some geometry to be much more sensitive to primordial non-Gaussian signals \cite{TheseKomatsu}.

\subsection{Weak lensing effects}

At the level of the temperature tri-spectra subtle differences can be observed depending on the nature of the potential high order correlation functions. This is then interesting to compare those results with the temperature trispectra
induce by weak lensing lensing effects. As mentioned in the introduction, this effect is the dominant low redshift second order coupling. For parity reason, one expects indeed the trispectrum - rather that the bispectrum - to acquire a non negligeable value from lensing effects \cite{CMBLens, TriLens, HuLens, HuTrispectre1}. Lensing effects amount to relate the observed temperature contrast, $\hat{\delta T}/T$, to the primordial one, $\delta T/T$, through
\be
\hT(\hg) = \frac{\delta T}{T}(\hg + \vdg) \;, 
\ee
where $\vdg$ is the lens induced displacement field. Usually, the displacement field is smaller than the angular scale under interest so that we may Taylor expand the temperature contrast as
\be
\frac{\delta T}{T}(\hg + \vdg) = \frac{\delta T}{T}(\hg) + \vdg \cdot \nabla  \frac{\delta T}{T}(\hg) + o( \vdg^2) \; .
\ee 
The first non vanishing contribution to the connected trispectrum is given by
\be
\label{trilens}
\langle \hT(\hg_1) \hT(\hg_2) \hT(\hg_3) \hT(\hg_4) \rangle_c  = \langle  \nabla_i \frac{\delta T}{T}(\hg_1) \frac{\delta T}{T}(\hg_3) \rangle \; \langle \nabla_j \frac{\delta T}{T}(\hg_2) \frac{\delta T}{T}(\hg_4) \rangle  \; \langle {\delta \gamma_1}_i \,  {\delta \gamma_2}_j \rangle + \mbox{11 other terms}\;, 
\ee 
where the Einstein index summation was used.
In the small angle approximation, the four-point function reads
\bea
\langle \hT(\hg_1) \hT(\hg_2) \hT(\hg_3) \hT(\hg_4) \rangle_c & =&  \int \frac{\dd^2 \ell_1}{2\pi} \frac{\dd^2 \ell_2}{2\pi} \frac{\dd^2 \ell_3}{2\pi} \frac{\dd^2 \ell_4}{2\pi} \; \delta_{\rm Dirac}(\vl_1+\vl_2+\vl_3+\vl_4)  {\rm e}^{i \vl_1 \cdot \hg_1 +i \vl_2 \cdot \hg_2 +i \vl_3 \cdot \hg_3 +i \vl_4 \cdot \hg_4 } \nonumber\\
&& \int \dd^2 L \; \left( C_{\ell_1} \vl_1 + C_{\ell_2} \vl_2 \right) \cdot \frac{\vec{L}}{L^2}  \; \times \; C_L^{\phi \phi}  \; \times \; \left( C_{\ell_3} \vl_3 + C_{\ell_4} \vl_4 \right) \cdot \frac{\vec{L}}{L^2} \; \delta_{\rm Dirac}(\vl_1+\vl_2-\vec{L}) \nonumber\\
&& + \left( 2\leftrightarrow 3\right) + \left( 2\leftrightarrow 4 \right) \;, 
\eea
where
\be
C_L^{\phi \phi} = - \frac{1}{(2\pi)^2} \; \int_0^{\chi_{\rm CMB}} \dd \chi \frac{w^2(\chi)}{{\cal D}_0^2(\chi)} P_{\delta} \left( \frac{L}{ {\cal D}_0(\chi)} \right) \; .
\ee
Here $\chi$ and ${\cal D}_0(\chi)$ are respectively the comoving distance and the comoving angular diameter distance from the observer while $P_{\delta}$ is the matter power spectrum. The function $w$ is a function of distances and strongly depends on the cosmological parameters (see e.g.~\cite{CMBLens, TriLens, HuLens}). 
We define the reduced trispectrum as
\bea
^{\rm lens} t^{\ell_1 \ell_2}_{\ell_3 \ell_4}(L) & =&  C_L^{\phi \phi} \left[ \left( C_{\ell_1} \vl_1 + C_{\ell_2} \vl_2 \right) \cdot \frac{\vec{L}}{L^2} \; \times \; \left[ C_{\ell_3} \vl_3 + C_{\ell_4} \vl_4 \right] \cdot \frac{\vec{L}}{L^2} \right] \nonumber\\
&&+ C_{L_{13}}^{\phi \phi} \left[ \left( C_{\ell_1} \vl_1 + C_{\ell_3} \vl_3 \right) \cdot \frac{\vec{L}_{13}}{L_{13}^2} \; \times \; \left[ C_{\ell_2} \vl_2 + C_{\ell_4} \vl_4 \right] \cdot \frac{\vec{L}_{13}}{L_{13}^2} \right] \nonumber\\
&&+  C_{L_{14}}^{\phi \phi} \left[ \left( C_{\ell_1} \vl_1 + C_{\ell_4} \vl_4 \right) \cdot \frac{\vec{L}_{14}}{L_{14}^2} \; \times \; \left[ C_{\ell_3} \vl_3 + C_{\ell_2} \vl_2 \right] \cdot \frac{\vec{L}_{14}}{L_{14}^2} \right] \;, 
\eea
and the normalized trispectrum reads
\be
^{\rm lens} {\tilde t}^{\ell_1 \ell_2}_{\ell_3 \ell_4}(L)=^{\rm lens}t^{\ell_1 \ell_2}_{\ell_3 \ell_4}(L) \frac{1}{\left( C_{\ell_1} + C_{\ell_2} \right) C_L \left( C_{\ell_3} + C_{\ell_4} \right)}.
\ee
The normalized trispectrum due to lensing effects is plotted in the right panels of Figs.~\ref{fig:TriEqui},~\ref{fig:Tri200} and~\ref{fig:Tri300} for different configurations in a sCDM model using the outputs of the CMBFast code~\cite{CMBFast}. As expected, the contribution of weak lensing occurs at small scales and globally increases with $\ell$. The signal is enhanced by large values of $L$ whose value triggs the positions of the acoustic peaks. 

Weak lensing effects have a specific signature on the temperature trispectrum which is quite different from the effects of primordial origin. The value of the trispectrum induced by weak lensing is much larger than the expected ones from primordial non-Gaussianities for $\nul \sim \nus \sim 1$. In particular, comparable signals in the range $\ell<50$ would require $\nu_3 \sim 10^{4}$ and even $\nu_3 \sim 10^{6}$ for $\ell>200$. 

The following sections aim at describing the general behaviors of correlation functions induced by primordial non-Gaussian statistics.

\section{Phenomenological approaches for the reconstruction of the correlation functions} 
 
\begin{figure} 
\centerline{ \epsfig {figure=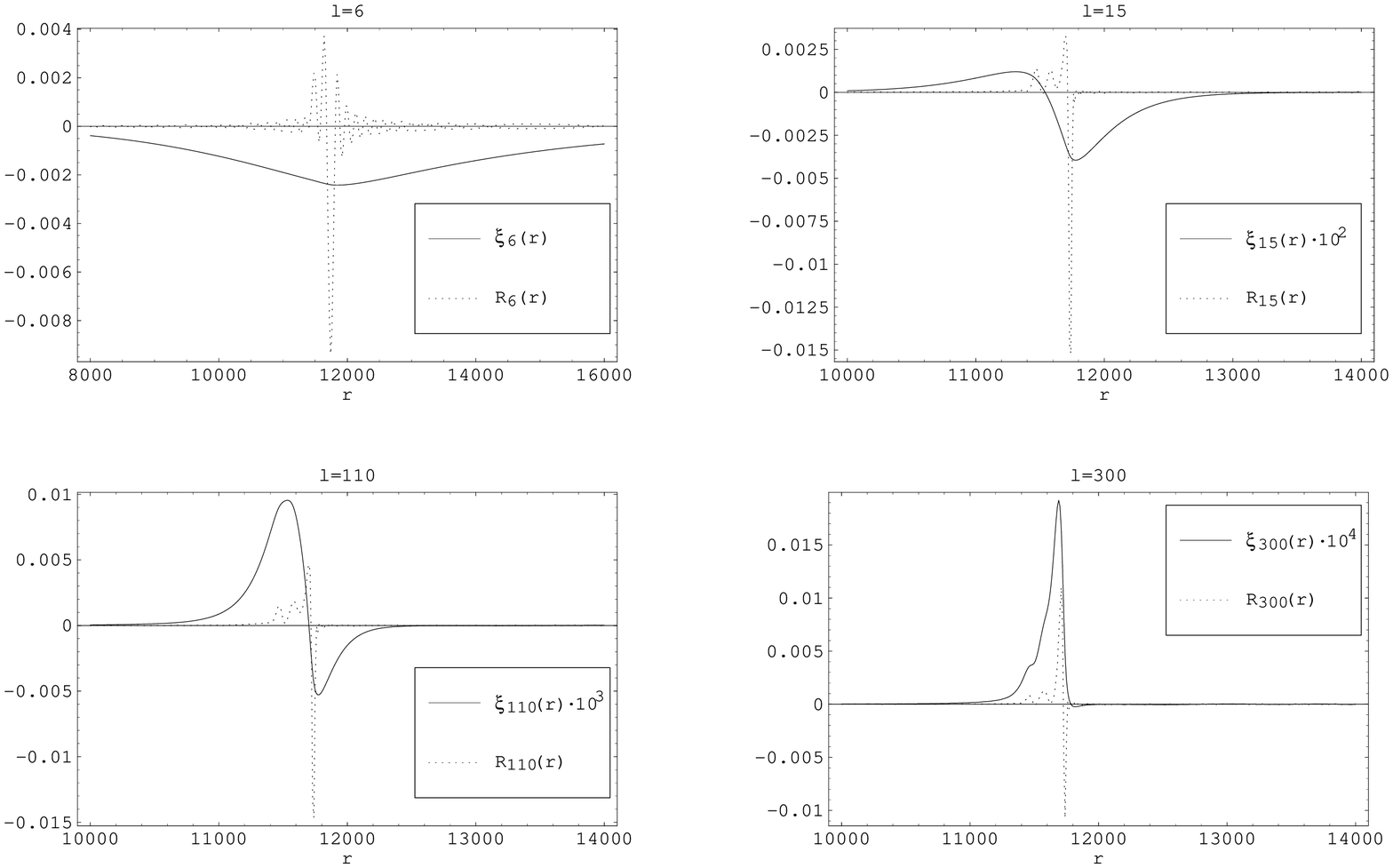,width=14cm}} 
\caption{Comparison between the variations of the radial transfer function $R_{\ell}(r)$ (dashed line) and $\xi_{\ell}(r)$ (solid line) as functions of $r$ for different $\ell$ in a sCDM model. From the upper left panel to the lower right panel: $\ell=6$, $\ell=50$, $\ell=110$, $\ell=300$.} 
\label{Rxifit} 
\end{figure}

As we saw in the previous sections, the correlation functions of the coefficients $a_{\ell m}$ are given in term of three functions: the radial transfer function $R_{\ell}(r)$, the function $\xi_{\ell}(r)$ and the vertex propagator $\zeta_{\ell}(r,r')$. Assuming a given power spectrum the propagator $\zeta_{\ell}(r,r')$ is a known function, peaked at $r = r'$ whereas $R_{\ell}(r)$ and $\xi_{\ell}(r)$ encode the details of the microphysics and have to be computed numerically. The  goal of this section is to describe the behaviors of these functions in a simple way.  As shown in Fig.~\ref{Rxifit}, the variations of $\xi_{\ell}(r)$ with $r$ are smoothed compared to those of $R_{\ell'}(r)$ which peaks at the last scattering surface $r=\rs$. Given the shape of the radial transfer function, it is then natural to expand it as a combination of a Dirac distribution, that would correspond to the Sachs Wolfe limit, and its first derivatives or some equivalently smoothed functions. This expansion seems valid as soon as one neglects the late integrated Sachs-Wolfe effect, which is not localized at the last scattering surface. This expansion reduces to a description of the temperature anisotropies today through the knowledge of the first radial derivatives of the gravitational potential on the last scattering surface: $\phi_{\ell m}(\rs)$,  $\phi'_{\ell m}(\rs)$, etc. This approach should be linked to a multipole expansion in the strong coupling limit. The goal of the next subsection is to test the validity of such an expansion.

\subsection{Which expansions for the transfer functions?} 
 
Neglecting the integrated Sachs-Wolfe effect and assuming a nearly instantaneous recombination, the description of the baryon-photon plasma dynamics is encoded by the first multipoles $\theta_{\ell}(k,\eta_*)$ at the last scattering surface. The transfer function can be approximately written with the following expansion
\be
\label{Tkexpand}
T_{\ell}(k) = \theta_0(k,\eta_*) j_{\ell}(k \rs) +  \theta_1(k,\eta_*) j'_{\ell}(k \rs) + ...\;, 
\ee
where, to keep insights of the micro-physics in mind, the monopole and the dipole are very roughly given by expressions such that~(see \cite{CMBSlow} for a more accurate phenomenological description of the acoustic pics),
\bea
\label{mono}
\theta_0(k,\eta_*) & \simeq & -\frac{1}{3} \cos(k c_s \eta_*) {\rm e}^{-k^2/k_D^2}\\
\label{dipo}
\theta_1(k,\eta_*) & \simeq & -\frac{1}{3\sqrt{3}} \sin(k c_s \eta_*) {\rm e}^{-k^2/k_D^2}\;. 
\eea
where the damping effect is schematically taken into account (and $k_D$ is the damping scale), $\eta_*$ is the conformal time at which recombination occurs, $c_s$ is the sound of speed of the plasma at the last scattering surface.

In the following we propose to replace the expansion (\ref{Tkexpand}), where the coefficients are $k$ dependent, by effective $l$-dependent coefficients, e.g., 
\be 
\label{Texpand} 
T_{\ell}(k)=A_0(\ell) j_{\ell}(k \rs) + A_1(\ell) j'_{\ell}(k \rs) + A_2(\ell) j''_{\ell}(k \rs) +... \; . 
\ee 
 This is justified from Eq. (\ref{Tkexpand}) by the fact that the Bessel functions, and their derivatives, peak at $k \sim \ell / \rs$. Roughly speaking we then expect that $A_{0}(\ell)\approx \theta_0(\ell/\rs,\eta_*)$, $A_{1}(\ell)\approx \theta_1(\ell/\rs,\eta_*)$. This will explain the rough behavior of these coefficients. In the following though we will not take this identification into account any more.

%

The idea is then, that at least for large enough angular scales, such an expansion could provide a reasonable description of the observed anisotropies when only a few terms are taken into account. Such a convergence is ensured for Eq.~(\ref{Tkexpand}) because the multipole decomposition at the last scattering surface is naturally ordered by a small parameter $k / \dot{\tau}$, where $\dot{\tau}$ is the recombination rate.  Incidentally one can remark that the first term of Eq.~(\ref{Texpand}) would correspond to the Sach-Wolfe effect in the limit where $A_0$ is independent on $\ell$.
In the following though, we treat these coefficients on a pure phenomenological footing. We will only assume that the decomposition~(\ref{Texpand}) is sensible and provides us with a good description of the transfer function.

Such an expansion implies the following form for the radial transfer function 
\be 
\label{Rexpand} 
R_{\ell}(r) = A_0(\ell) g_0(r,\rs;\ell) + A_1(\ell) g_1(r,\rs;\ell) +  A_2(\ell) g_2(r,\rs;\ell) +...\;, 
\ee 
with 
\be
\label{gn}
g_n(r,\rs ; \ell) = \int \dd k \; k^2 r^2 \; j_{\ell}(k r) j^{(n)}_{\ell}(k \rs) \; .
\ee 
 
\begin{figure} 
\centerline{ \epsfig {figure=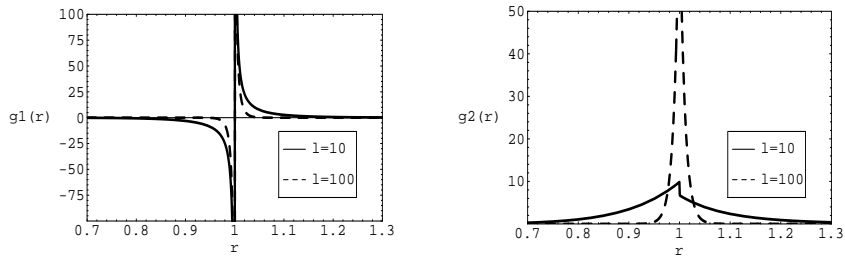,width=12cm,height=4cm}} 
\caption{Shape of $g_1$ (left panel) and $g_2$ (right panel) as a function of $r/\rs$ for $\ell=10$ (solid line) and $\ell=100$ (dashed line).} 
\label{delta} 
\end{figure}

The $g_1$ function is given in terms of the hypergeometric function $_2 \! F_1$ and corresponds to a $\ell$-dependent smoothing of $\delta_{\rm Dirac}'(r-\rs)$. The expressions of $g_1$ and $g_2$ are given in appendix~\ref{AppExpansion} and their shapes are shown for different $\ell$ in Fig.~\ref{delta}. Note that $g_0(r,\rs;\ell)= \delta_{\rm Dirac}(r-\rs) \pi / 2\rs^2$.

To describe the effect of the expansion~(\ref{Texpand}), it is necessary to know the coefficients $A_i(\ell)$. We then adopt a phenomenological approach in which these coefficients are determined from a comparison with the CMBFast code results~\cite{CMBFast}. All the following results are obtained in a standard cold dark matter (sCDM) model, where the integrated effect may be neglected.

\begin{figure} 
\centerline{
\begin{tabular}{ccc} \epsfig {figure=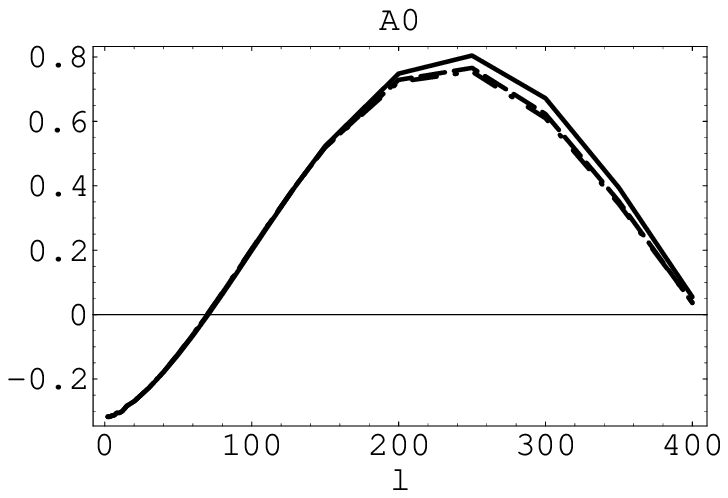, width=6cm,height=4cm} &  \epsfig {figure=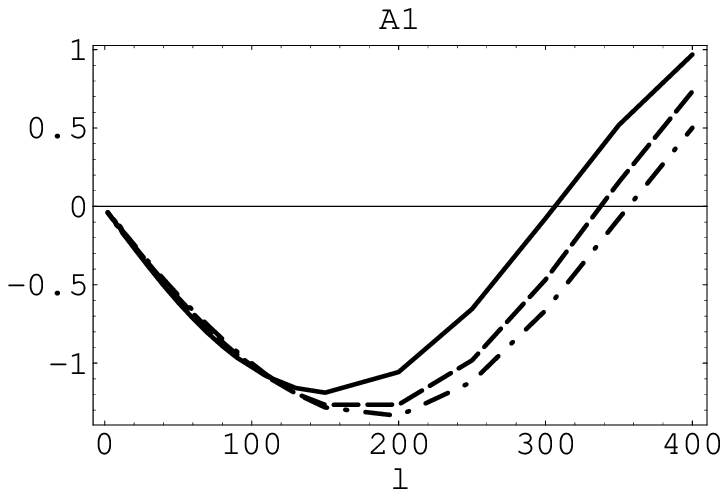, width=6cm,height=4cm} & \epsfig {figure=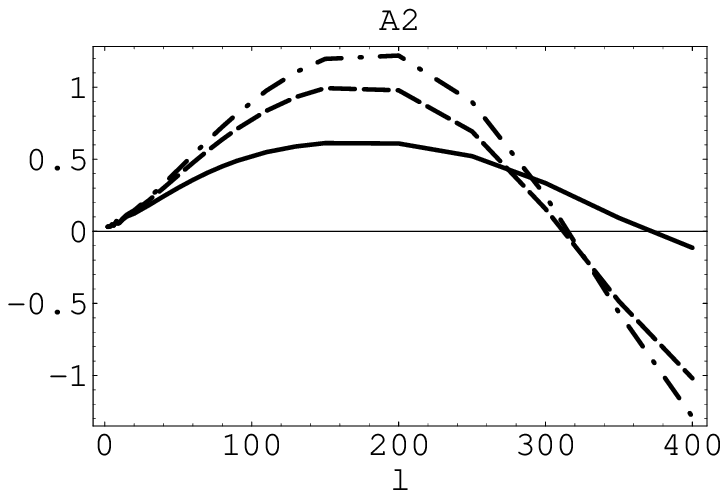, width=6cm,height=4cm}
\end{tabular} } 
\caption{Shapes of the coefficients $A_0$ (left panel), $A_1$ (center panel) and $A_2$ (right panel) as functions of $\ell$ in a sCDM model. These coefficients have been obtained from a fit of the standard CDM outputs of the CMBFast code with the ansatz Eq.~(\ref{xiexpand}) for different choice of $\rs$. Solid line: $\rs=11670 \;{\rm h}^{-1}$ Mpc. Dashed line: $\rs=11690 \; {\rm h}^{-1}$ Mpc. Dotted dashed line: $\rs=11700\;  {\rm h}^{-1}$ Mpc.} 
\label{Ajcomp} 
\end{figure}

\begin{figure} 
\centerline{
\begin{tabular}{cc} \epsfig {figure=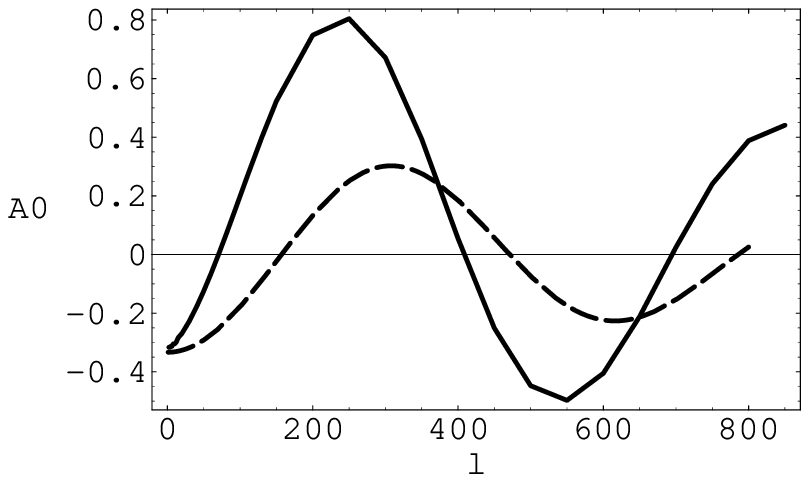, width=7cm,height=4cm} & \epsfig {figure=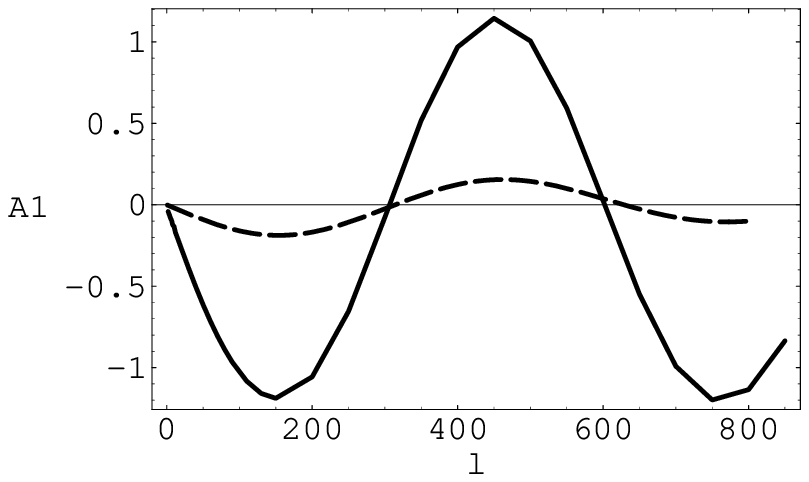, width=7cm,height=4cm} 
\end{tabular} } 
\caption{Comparison between the fitted coefficients (solide line) $A_0(\ell)$ (left panel) and $A_1(\ell)$ (right panel) and the roughly estimated expected ones (dashed line) from Eqs.~(\ref{mono}) and (\ref{dipo}) with the typical values $c_s \sim 10^{-1}$, $\rs / \eta_*\sim 10^3$ and $k_D \rs \sim 10^3$. Eqs.~(\ref{mono}) and (\ref{dipo}) give a rough estimate of the coefficients $A_0$ and $A_1$ and explain their oscillations with $\ell$. } 
\label{monodipo} 
\end{figure}

In principle, one should be capable of extracting the coefficients $A_0$, $A_1$, $A_2$,... by fitting the numerical radial transfer function with a combination of the functions $g_0$, $g_1$, $g_2$,... . However, this approach raises some issues. In particular, the determination of the coefficients $A_i$ is extremely sensitive to the choice of $\rs$ because the functions $g_i$ peak at the last scattering surface. On the other hand, the expansion~(\ref{Texpand}) induces the following expansion of the smoothed functions $\xi_{\ell}(r)$: 
\be
\label{xiexpand} 
\xi_{\ell}(r) = A_0(\ell) f_0(\frac{r}{\rs};\ell) + A_1(\ell) f_1(\frac{r}{\rs};\ell) + A_2(\ell) f_2(\frac{r}{\rs};\ell) +...\;, 
\ee 
where, for a scale-invariant power spectrum,  
\be
\label{fn}
f_n(\frac{r}{\rs};\ell)  =  \int \frac{\dd k}{k} \; j_{\ell}(k r) j^{(n)}_{\ell}(k \rs)  \; .
\ee
The functions $f_i$ are computed for $i=0,1,2$ in appendix~\ref{Appbis} and plotted in Fig.~\ref{f012}.
\begin{figure} 
\centerline{ \epsfig {figure=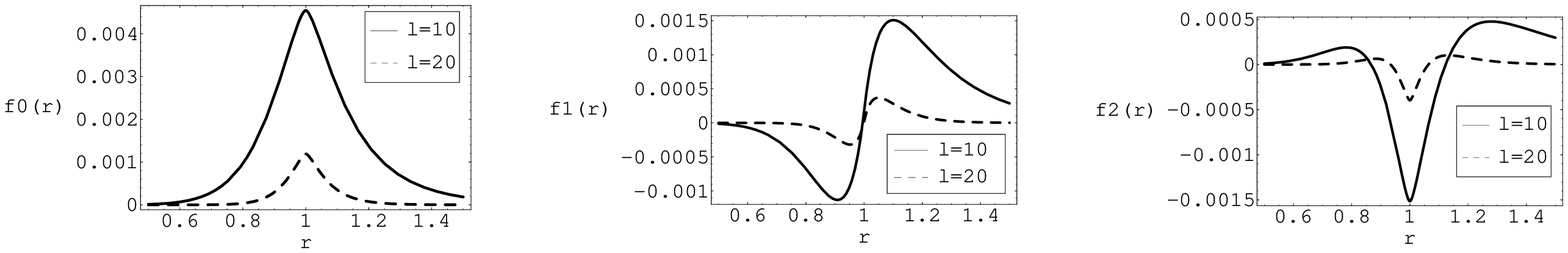, width=18cm,height=4cm}} 
\caption{Shape of $f_0$ (left panel), $f_1$ (center panel) and $f_2$ (right panel) as a function of $r/\rs$ for $\ell=10$ (solid line) and $\ell=20$ (dashed line) (sCDM model).} 
\label{f012} 
\end{figure} 
Fitting the functions $\xi_{\ell}(r)$ with a combination of the functions $f_0$, $f_1$ and $f_2$ gives an accurate determination of the coefficients $A_0$, $A_1$ and $A_2$. The results are shown in Fig.~\ref{Ajcomp}. Comparisons with the expected coefficients $A_0$ and $A_1$ in Eqs.~(\ref{mono}) and (\ref{dipo}) for typical values of $c_s$ and $k_D$ are shown in Fig.~\ref{monodipo}. We can see that the oscillations of the coefficients originate from the oscillations of the first multipoles on the last scattering surface. As expected from the Sachs-Wolfe limit, $A_0(\ell)$ reduces to the constant value $\sim -1/3$ at low $\ell$ and $A_1(\ell),A_2(\ell) \rightarrow 0$ when $\ell$ goes to zero. Note that neglecting the early integrated Sachs-Wolfe effect in Eq.~(\ref{Tkexpand}) for a sCDM model do not modify much the expected coefficient $A_0$ at low multipole values $\ell$. This would not be the case in a $\Lambda$CDM model for instance, where the integrated Sachs-Wolfe effect would induce large values of $A_0$ at low $\ell$.

\subsection{Power spectrum} 
 
Given the coefficients $A_p$, the computation of the $C_\ell$'s may be performed analytically. Integrating over the radial coordinate $r$ in Eq.~(\ref{Clexpr}) yields
\be
C_{\ell}  = \frac{2}{\pi} \int \frac{{\rm d}k}{k} T_{\ell}(k)^2 \;, 
\ee
where we set a scale invariant power spectrum.
Using the expansion~(\ref{Texpand}) leads to
\bea
\label{Cltot} 
C_{\ell} &=& \frac{2}{ \pi } \Bigg[ \frac{A_0^2}{2 \ell (\ell +1)} +\frac{A_1^2}{6 (\ell-1)(\ell+2)}+ A_2^2 \frac{3 \ell^2+3\ell-10}{30\ell(\ell-2)(\ell+1)(\ell+3)}   - \frac{A_0 A_2}{3 \ell(\ell+1)} \nonumber \\ 
&&+ A_0 A_1 \frac{\pi (2\ell-3)!! }{(2 \ell +3 )!!}  +  A_1 A_2  \frac{ \pi (\ell^2+\ell+3)(2\ell-5)!!}{(2\ell+5)!!}  + ... \Bigg]\;, 
\eea 
where higher order terms were not written explicitly as they are expected to be small corrections to this result. From the above expression, we can see that all the terms involving $A_i A_j$ are subdominant when $i+j$ is odd. This is a quite general fact as we will see in the flat sky approximation (see appendix~\ref{sec:flatsky}). For sufficiently large $\ell$, only the dominant terms contribute and the two-point function becomes 
\be 
\label{Clapprox} 
C_{\ell} =  \frac{2}{ \pi \ell^2 } \left[ \frac{A_0^2}{2} + \frac{A_1^2}{6}+\frac{A_2^2}{10} - \frac{A_0 A_2}{3} + ...\right] \; . 
\ee 

\begin{figure} 
\centerline{ \epsfig {figure=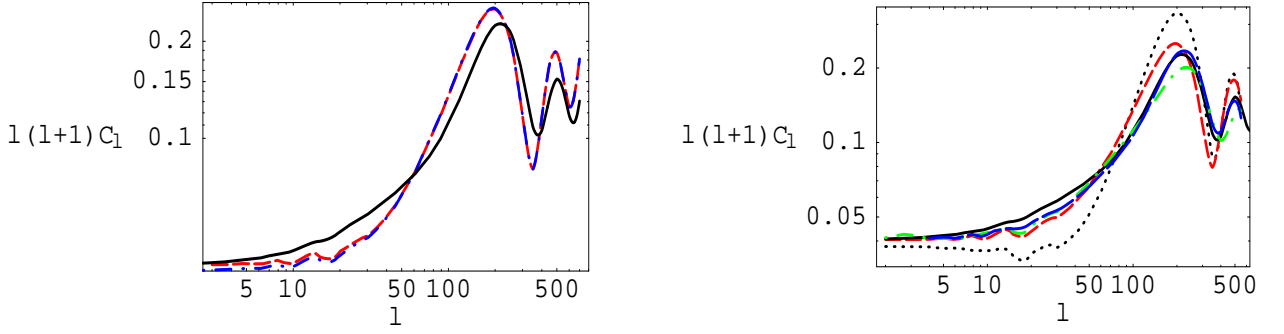, width=18cm,height=6cm}} 
\caption{Reconstruction of the power spectra from the ansatz (\ref{Texpand}). The coefficient $A_i$ are obtained from a fit of the standard CDM outputs of the CMBFast code. Left panel: effect of the flat sky approximation on the power spectrum recontruction. The solid black line shows the results from CMBFast. The red dashed line shows the results from Eq.~(\ref{Cltot}) whereas the blue dotted-dashed line is obtained from the flat sky approximation Eq.~(\ref{Clapprox}). Those two recontructed curves are in agreement with each other for $\ell > 15$. Right panel: reconstruction of the power spectrum with different degrees of approximation: expansion to first order (contribution of $A_0$ and $A_1$ only) (dotted black line), to second order (long-dashed red line), to third order (dotted dashed green line) and to fourth order (short-dashed blue line). The solid black line represents the expected result from the CMBFast code.} 
\label{Cl_fit_tot} 
\end{figure}

Using the values of the previously obtained coefficients $A_i$, Eq.~(\ref{Clapprox}) appears to be a rough estimate of the power spectrum as shown in Fig.~\ref{Cl_fit_tot}. 

As we have shown in the previous subsection, the computations of higher order correlation functions involve quite cumbersome integrals over hypergeometric functions, which cannot be expressed analytically. To get an approximate result, it would seem natural to expand $\xi_{\ell}(r)$ about $r=\rs$ with a Taylor expansion since its behavior is smoothed compared to the radial transfer function's. However this series expansion appears not to be valid in a sufficiently wide region around $r= \rs$.  
As this approach does not allow to compute high order correlation functions, we should use another approximation. For sufficiently large $\ell$ ($\ell > 10$ in practice), Fig.~\ref{Cl_fit_tot} shows that the two-point function is well described only by the leading order terms in $\ell$. As the large $\ell$ approximation corresponds to small angular scales, it also means that the sky is assumed to be locally flat so that projection effects reduce to a simple identification. In the following we assume that the last scattering surface is properly described by a plan. Hence, the multipole decomposition corresponds to a flat two-dimensional Fourier transform. 
 
Using the expression of the temperature contrast 
\be 
\frac{\delta T}{T} (\hg) = \sum_{\ell, m} a_{\ell  m} Y_{\ell m} \;, 
\ee 
we get 
\be 
\frac{\delta T}{T} (\hg) = \sum_{\ell, m} 4 \pi  (-i)^\ell \int \frac{\dd ^3 k}{(2 \pi)^{3/2}} T_{\ell}(k) \Phi(\vk) Y_{\ell m}(\hg)  Y^*_{\ell m}(\hk) \; . 
\ee 
 
The decomposition of the transfer function in terms of Bessel functions gives
\be 
\frac{\delta T}{T} (\hg) =  \sum_p \sum_{\ell, m} A_p(\ell) 4 \pi (-i)^\ell \int \frac{\dd ^3 k}{(2 \pi)^{3/2}} j^{(p)}_{\ell}(k \rs) \Phi(\vk) Y_{\ell m}(\hg)  Y^*_{\ell m}(\hk) \; . 
\ee 
Considering that the functions $A_p(\ell)$ do not vary much compared to the oscillating Bessel functions, we can use Eq.~(\ref{expbes}) to get 
\be 
4 \pi \sum_{\ell, m} (-i)^\ell j^{(p)}_{\ell}(k \rs) Y_{\ell m}(\hg)  Y^*_{\ell m}(\hk) = \frac{1}{k^p} \frac{\dd ^p}{\dd \rs^p} \; {\rm e}^{i \vk \cdot \vec{\rs}} \;, 
\ee 
with 
\bea 
\vk &=& k_z \vec{u}_z + \vk_{\perp} \;, \\ 
\vec{\rs} &=& \rs \hg \;, \\
\vk \cdot \vec{\rs} & \simeq & k_z \rs + \vl \cdot \hg \; .
\eea 
The $z$-axis, whose unit vector $\vec{u}_z$ is oriented towards us, supports the line of sight. Vectorial quantities belonging to the plan which is orthogonal to the line of sight are indexed by a symbol $\perp$. The vector $\vl=\vk_{\perp} / \rs$ is defined as the conjugated variable of the angle $\hg$. 

The temperature contrast becomes 
\bea 
\frac{\delta T}{T} (\hg) &=& \sum_p \int \frac{\dd^3 k}{(2 \pi)^{3/2}} A_p  \Phi(\vk) \, \frac{\dd^p}{k^p \, \dd \rs^p } \; {\rm e}^{-i \vk \cdot \vec{\rs}}\\
&=&  \frac{1}{(2\pi)^{3/2} \, \rs^3} \sum_p \int \dd x \, \dd^2 \vl \; A_p \; \frac{(-i)^p \, x^p \; {\rm e}^{-ix}}{(x^2+\ell^2)^{p/2}} \; \Phi(\vk) {\rm e}^{-i \vl \cdot \hg}\;, \; \; \mbox{with} \; \; \vk=\frac{x}{\rs}\vec{u}_z+\frac{\vl}{\rs} \; . 
\eea 
and the two-point function reads 
\be 
\langle \frac{\delta T}{T} (\hg_1) \frac{\delta T}{T} (\hg_2) \rangle = \sum_{p_1,p_2} \int \frac{\dd^2 \vl}{(2\pi)^2} F_{p_1,p_2}(\ell) A_{p_1}(\ell) A_{p_2}(\ell) {\rm e}^{-i \vl \cdot (\hg_1-\hg_2)}\;, 
\ee 
where
\be
F_{p_1,p_2}(\ell) = (-i)^{p_1-p_2} \int \frac{\dd x}{2\pi} \; \tilde{P}\left(\sqrt{x^2+\ell^2}\right) \frac{x^{p_1+p_2}}{(x^2+\ell^2)^{(3+p_1+p_2)/2}}\; \;  \mbox{with} \; \; \tilde{P}(y)=\rs^3 P\left(\frac{y}{\rs}\right).
\ee 
 
For a scale invariant power spectrum $\tilde{P}(y)=1/y^3$, the following integrals
\be 
\int_{-\infty}^{+\infty} \dd x \frac{x^{p_1+p_2}}{(x^2+\ell^2)^{(3+p_1+p_2)/2}} = \left\{ \begin{array}{cc} 0 & \mbox{if $p_1+p_2$ is odd} \\ 
\\  
\frac{2}{1+p_1+p_2} \frac{1}{\ell^2} & \mbox{if $p_1+p_2$ is even} \end{array} \right. \nonumber 
\ee 
lead to
\be 
\langle \frac{\delta T}{T} (\hg_1) \frac{\delta T}{T} (\hg_2) \rangle = \frac{2}{\pi} \int \frac{\dd^2 \vec{\ell} }{(2 \pi)^2} \; \frac{1}{\ell^2} \; \left[ \frac{A_0^2}{2} + \frac{A_1^2}{6}+\frac{A_2^2}{10} +\frac{A_3^2}{14}  +\frac{A_4^2}{18} - \frac{A_0 A_2}{3} +  \frac{A_0 A_4}{5} - \frac{A_1 A_3}{5} -  \frac{A_2 A_4}{7} + ...\right] \; {\rm e}^{-i \vl \cdot (\hg_1-\hg_2)} \;, 
\ee 
which is in perfect agreement with Eq.~(\ref{Clapprox}) (see appendix~\ref{sec:flatsky} for the correspondence between flat sky and all sky formalisms). As we highlighted, only terms $A_i A_j$ with $i+j$ even do contribute to the two point function. The convergence of the series may be checked in Fig.~\ref{Cl_fit_tot} where we used the expansion~(\ref{Texpand}) to $A_4$.

The large $\ell$ limit together with the expansion~(\ref{Texpand}) provides us with a more tractable formalism to express higher order correlation functions. In the following, we explore the approximate behavior of the bispectrum through the very first terms of the expansion~(\ref{Texpand}).  

\subsection{Bispectrum} 
 
An accurate description of the bispectrum is a powerful tool towards discriminating among the inflationary models~\cite{NGShape, NGSignatures} or understanding the primary or secondary nature of non-Gaussianities~\cite{NGTemp, SNforNG, bispectreform, Aniso}. In this perspective, we apply our expansion to explore the key features of the temperature three-point function which may be expressed as
\be 
\langle \frac{\delta T}{T} (\hg_1) \frac{\delta T}{T} (\hg_2)  \frac{\delta T}{T} (\hg_3) \rangle_c = \sum_{p_1,p_2,p_3} \int \frac{\dd^2 \vl_1}{2 \pi}\frac{\dd^2 \vl_2}{2 \pi}\frac{\dd^2 \vl_3}{2 \pi} \;  A_{p_1}  A_{p_2}  A_{p_3} \; F_{p_1,p_2,p_3}(\ell_1,\ell_2,\ell_3) \; {\rm e}^{-i \vl_1 \cdot \hg_1}  {\rm e}^{ -i \vl_2 \cdot \hg_2}{\rm e}^{ -i \vl_3 \cdot \hg_3} \delta_{\rm Dirac} \left( \vl_1 + \vl_2 + \vl_3 \right)
\ee
with
\bea
F_{p_1,p_2,p_3}(\ell_1,\ell_2,\ell_3) &=& \nu_2 \, \frac{(-i)^{p_1+p_2+p_3}}{(2\pi)^{3/2}} \int \dd x_1 \dd x_2 \dd x_3 \;  \delta_{\rm Dirac} \left( x_1 + x_2 + x_3 \right) \nonumber\\
&&\left[ \frac{\tilde{P}(\sqrt{x_1^2+\ell_1^2}) \, x_1^{p_1} \, {\rm e}^{-ix_1}}{(x_1^2+\ell_1^2)^{p_1/2}} \; \frac{\tilde{P}(\sqrt{x_2^2+\ell_2^2}) \, x_2^{p_2} \, {\rm e}^{-ix_2}}{(x_2^2+\ell_2^2)^{p_2/2}} \; \frac{x_3^{p_3} \, {\rm e}^{-ix_3}}{(x_3^2+\ell_3^2)^{p_3/2}} +{\rm perm.} \right]\; . 
\eea 
 
From the changes of variables $x_i \rightarrow -x_i$, one can see that the non-vanishing terms are those with $p_1+p_2+p_3$ even. In the special case where $p_3=0$, both $p_1$ and $p_2$ should be even.  
 
As the direct computation is uneasy, we decompose the Dirac function $ \delta_{\rm Dirac} \left( x_1 + x_2 + x_3 \right)$ as in Eq.~(\ref{diracexp}) to recast the three-point function into 
\be 
\label{T3flat}
\langle \frac{\delta T}{T} (\hg_1) \frac{\delta T}{T} (\hg_2)  \frac{\delta T}{T} (\hg_3) \rangle_c = \frac{\nu_2}{(2 \pi)^{3/2}} \int \frac{\dd^2 \vl_1}{2 \pi} \int \frac{\dd^2 \vl_2}{2 \pi} \int \frac{\dd^2 \vl_3}{2 \pi} \;  \delta_{\rm Dirac} \left( \vl_1 + \vl_2 + \vl_3 \right) {\rm e}^{-i \left(\hg_1 \cdot \vl_1+\hg_2 \cdot \vl_2+\hg_3 \cdot \vl_3\right)}b_{(\ell_1, \ell_2, \ell_3)} \;, 
\ee 
with
\be
b_{(\ell_1, \ell_2, \ell_3)} = {\cal B}_{(\ell_1, \ell_2, \ell_3) } + {\cal B}_{(\ell_3, \ell_1, \ell_2) }  + {\cal B}_{(\ell_2, \ell_3, \ell_1) } \; .
\ee
For a scale invariant power spectrum
\bea
\label{Biexpand} 
{\cal B}_{(\ell_1, \ell_2, \ell_3) } &=& \frac{4}{\ell_1^2 \ell_2^2}  A_0(\ell_1) A_0(\ell_2) A_0(\ell_3) \nonumber \\ 
&&+ 2 \frac{\ell_3}{\ell_1 \ell_2} b_1\left(\ell_1, \ell_2, \ell_3 \right) A_1(\ell_1) A_0(\ell_2) A_1(\ell_3) + (\ell_1 \leftrightarrow \ell_2) \nonumber \\ 
&&+ 4 \frac{\ell_3}{\ell_1 \ell_2} b_1\left(\ell_3, \ell_1, \ell_2 \right) A_0(\ell_1) A_0(\ell_2) A_2(\ell_3) \nonumber \\ 
&& - \frac{4}{3 \ell_1^2 \ell_2^2}  A_2(\ell_1) A_0(\ell_2) A_0(\ell_3) + (\ell_1 \leftrightarrow \ell_2) \nonumber \\ 
&&+ \frac{\pi^2}{2} \; \frac{\ell_3}{\ell_1 \ell_2 (\ell_1+\ell_2+\ell_3)^3}  A_1(\ell_1) A_1(\ell_2) A_2(\ell_3)  \nonumber \\ 
&&-\frac{2}{3} \;  \frac{\ell_3}{\ell_1^2 \ell_2}  b_2 \left(\ell_1, \ell_2, \ell_3 \right) A_2(\ell_1) A_1(\ell_2) A_1(\ell_3) +( \ell_1 \leftrightarrow \ell_2) \nonumber \\ 
&& +\frac{4}{9} \; \frac{1}{\ell_1^2 \ell_2^2} A_2(\ell_1) A_2(\ell_2) A_0(\ell_3) \nonumber \\ 
&& -\frac{4}{3} \; \frac{\ell_3}{\ell_1^2 \ell_2} b_2\left(\ell_1, \ell_3, \ell_2 \right) A_2(\ell_1) A_0(\ell_2) A_2(\ell_3) + (\ell_1 \leftrightarrow \ell_2) \nonumber\\
&&+\frac{4}{9} \, \frac{\ell_3}{\ell_1^2 \ell_2^2} b_3\left(\ell_1, \ell_2, \ell_3 \right) A_2(\ell_1) A_2(\ell_2) A_2(\ell_3) \; .    
\eea 
We defined the following functions (see appendix~\ref{Appbis})
\bea 
b_1 \left( \ell_1, \ell_2, \ell_3 \right) &=& \int_{0}^{\infty} \dd z \; z^2 \; {\rm e}^{-\ell_1 z}\;  K_1( \ell_2 z) \; K_1( \ell_3 z) \\ 
b_2 \left( \ell_1, \ell_2, \ell_3 \right) &=& \int_{0}^{\infty} \dd z \; z \;{\rm e}^{- \ell_2 z}\; K_1(\ell_3 z) \; \left[ \ell_1 z K_1(\ell_1 z)-\ell_1^2 z^2 K_0(\ell_1 z) \right] \\
b_3\left( \ell_1, \ell_2, \ell_3 \right) &=& \int_{0}^{\infty} \dd z \;{\rm e}^{- \ell_3 z}\; \left[ \ell_1 z K_1(\ell_1 z)-\ell_1^2 z^2 K_0(\ell_1 z) \right]   \; \left[ \ell_2 z K_1(\ell_2 z)-\ell_2^2 z^2 K_0(\ell_2 z) \right] \;, 
\eea 
where $K_n$ stands for the modified Bessel function of the second kind of order $n$.

The link between the flat sky and all sky formalisms imposes (see appendix~\ref{sec:flatsky})
\be
b_{\ell_1 \ell_2 \ell_3 } = \frac{1}{\sqrt{2 \pi}}  b_{(\ell_1,\ell_2,\ell_3)} \; .
\ee 

In the peculiar case where $\ell_1=\ell_2=\ell_3=\ell$, the functions $b_1$, $b_2$ and $b_3$ simplify to 
\bea 
b_1(\ell, \ell, \ell ) &=& \frac{1}{\ell^3} \left[ \frac{3 \pi^2}{32} \; {_2} \! F_1 \left( \frac{3}{2}, \frac{5}{2}; 2; \frac{1}{4} \right) - \frac{2}{3} {_3} \! F_2 \left( 1,2,3 ; \frac{3}{2}, \frac{5}{2} ; \frac{1}{4} \right) \right] \simeq \frac{0.526}{\ell^3} \\ 
b_2(\ell, \ell, \ell ) &\simeq & 0.381 \frac{1}{\ell^2}\\
b_3(\ell, \ell, \ell ) &\simeq & 0.335 \frac{1}{\ell} \; . 
\eea  

\begin{figure} 
\centerline{ \epsfig {figure=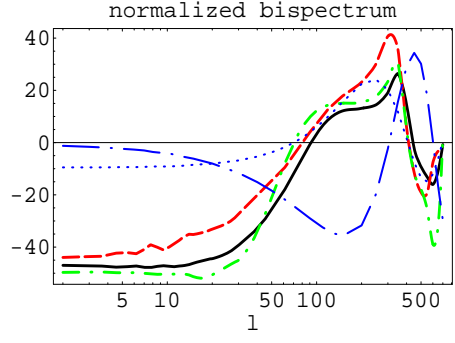, width=7cm,height=4.5cm}} 
\caption{Normalized bispectrum for an equilateral 
configuration $\tilde{b}_{\ell \ell \ell}$ in a sCDM 
model with $\nu_2=1$. The solid black line is the expected bispectrum. The red dashed and the green dot-dashed lines are respectively the reconstructed bispectrum to first and second order. The blue lines show the behavior of the coefficients $A_0(\ell)$ (dotted line) and $A_1(\ell)$ (long-dash-dotted line).} 
\label{fig:biequi} 
\end{figure} 

\begin{figure}
\centerline{ 
\epsfig {figure=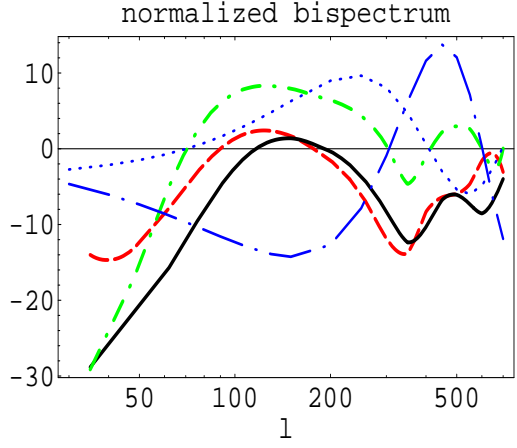, width=8cm,height=6cm } }
\caption{Reconstruction of the normalized bispectrum 
for an isosceles configuration  $\tilde{b}_{\ell \, \ell \, 70}$ 
as a function of $\ell$ in a sCDM model with $\nu_2=1$. 
The solide black line represents the expected 
bispectrum. The green dotted dashed line and 
the red dashed line represent the reconstructed 
bispectrum from an expansion to the order $A_1$ 
and $A_2$ respectively. The blue lines represents the coefficients $A_0(\ell)$ (dotted line) and 
$A_1(\ell)$ (long dashed dotted line).} 
\label{fig:biso} 
\end{figure}

\begin{figure}
\centerline{
\epsfig {figure=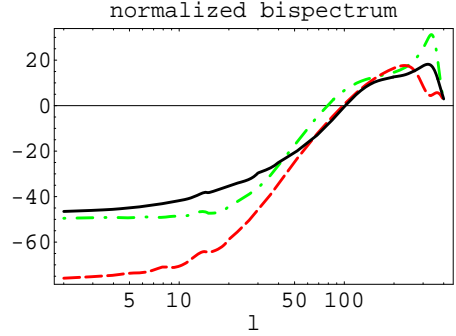, width=7cm,height=4.5cm}  } 
\caption{Normalized bispectrum for an isosceles configuration $\tilde{b}_{\ell \, \ell+300 \, \ell+300}$(left panel) in a sCDM model (solid black line) with $\nu_2=1$. The green dotted dashed line and the red dashed line represent the reconstructed normalized bi-spectrum to order $A_1(\ell)$ and $A_2(\ell)$ respectively. The blue lines represent the coefficients $A_0$ (dotted line) and $A_1$ (long-dash-dotted line).}
\label{fig:bi300plus} 
\end{figure}

Numerical results are shown in Figs.~\ref{fig:biequi},~\ref{fig:biso} and ~\ref{fig:bi300plus}. In Fig.~\ref{fig:biequi}, the normalized bispectrum in~(\ref{T3flat}) is compared to the expected one in an equilateral configuration. For an equilateral configuration, it appears that the bispectrum is properly approximated with a few terms. In particular, one can see that the main features of the bispectrum are encoded by the coefficients $A_0$ and $A_1$. Recalling Eq.~(\ref{Biexpand}), it means that roughly $b_{\ell \ell \ell}\propto A_0(\ell) (A_0^2(\ell)+ 0.5 A_1^2(\ell))$. Hence the zeros of the bispectrum are roughly given by the zeros of the monopole $A_0(\ell)$ for an equilateral configuration (see Fig.~\ref{fig:biequi}).

Fig.\ref{fig:biso} represents the peculiar configuration where one of the $\ell$ is fixed to a zero of the monopole, $\ell=70$, while the other two are equal. From Eq.~(\ref{Biexpand}), the bispectrum should reduce to first order to a single term
\be
b_{ 70\, \ell\, \ell} \propto A_0(\ell) A_1(\ell) A_1(70).
\ee 
Hence the zeros of the bispectrum are roughly those of the monopole and the dipole terms. Plots of Fig.\ref{fig:biso} shows that this approximation is too crude to estimate the zeros of the bispectrum mainly because this configuration demands the dominant term to vanish. However the coarse features are recovered and the sub-dominant terms contribute as a global shift that lead to a proper fit of the expected bispectrum.

We also paid attention to configurations that would correlate the first two acoustic peaks of the $C_\ell$'s. Fig.~\ref{fig:bi300plus} shows the bispectrum $b_{\ell \, \ell+300 \, \ell+300}$. The basic features of this bispectrum may be infered by noting that $A_0(\ell+300) \simeq -A_0(\ell)$ and $A_1(\ell+300) \simeq -A_1(\ell)$. We then get
\be
b_{\ell \, \ell+300 \, \ell+300} \propto A_0(\ell) \left[ A_0(\ell)^2 + .5 A_1(\ell)^2 \right] \; \; \; \mbox{and} \; \; \; b_{\ell \, \ell \, \ell+300} \propto - A_0(\ell) \left[ A_0(\ell)^2 + h(\ell) A_1(\ell)^2 \right]\; ,
\ee  
where $h(\ell) \sim 0.25$ is a slowly varying function. These configurations roughly reproduce the bispectrum corresponding to an equilateral configuration.

Our results provide a simple way towards understanding the behavior of the bispectrum as a function of the configuration. Recalling that the coefficients $A_0$, $A_1$ and $A_2$ represent the first momenta of the gravitational potential on the last scattering surface, we have shown that the coarse features of the bispectrum may be infered from the monopole and the dipole behaviors.

\subsection{Trispectrum}

The analysis of the trispectrum may be performed in the same way as previously for the bispectrum. As the number of terms in our phenomenological approach increases with the order of the correlation function, the key features of the trispectrum become less obvious to infer. However the sign of the trispectrum may be roughly predicted given the behaviors of $A_0$ and $A_1$.

\begin{figure}
\centerline{ 
\epsfig {figure=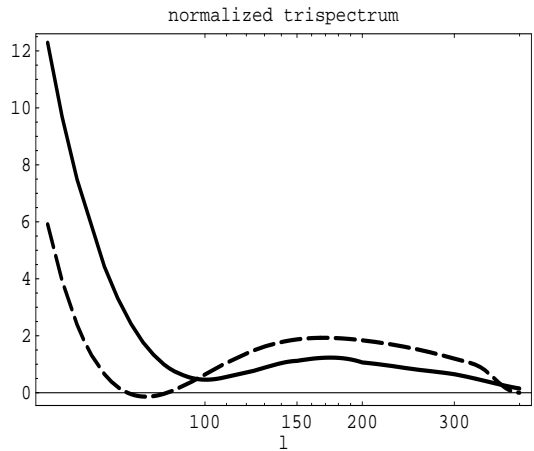, width=8cm,height=6cm } }
\caption{Plot of $^{\rm star} {\cal T}^{\ell \; \ell}_{\ell+300 \, \ell+300}$(L) as a function of $\ell$ for $L=5$ with $\nus=1$. The solid line correspond to the CMBFast code results whereas the dashed one is the reconstructed trispectrum.} 
\label{tstar300Approx} 
\end{figure}

The very first terms of the expansion of the trispectrum read
\bea 
 \langle \frac{\delta T}{T} (\hg_1) \frac{\delta T}{T} (\hg_2)  \frac{\delta T}{T} (\hg_3)  \frac{\delta T}{T} (\hg_4) \rangle_c &=& \frac{1}{(2 \pi)^2} \int \frac{\dd^2 \vl_1}{2 \pi} \int \frac{\dd^2 \vl_2}{2 \pi} \int \frac{\dd^2 \vl_3}{2 \pi}  \int \frac{\dd^2 \vl_4}{2 \pi} \;  \delta_{\rm Dirac} \left( \vl_1 + \vl_2 + \vl_3 + \vl_4 \right) {\rm e}^{-i \left(\hg_1 \cdot \vl_1+\hg_2 \cdot \vl_2+\hg_3 \cdot \vl_3+\hg_4 \cdot \vl_4 \right)} \nonumber\\
&& \int \dd^2 \vec{L} \left[ \delta_{\rm Dirac} \left(\vl_1+\vl_2+\vec{L} \right) t^{(\ell_1, \ell_2)}_{(\ell_3, \ell_4)}(L) +\left( 2 \leftrightarrow 3 \right) +\left( 2 \leftrightarrow 4 \right) \right] \; , 
\eea
with
\bea
^{\rm star}t^{\ell_1 \ell_2}_{\ell_3 \ell_4}(L) &=& \frac{8 \nus}{ 3 \ell_1^2 \ell_2^2 \ell_3^2} \; A_0(\ell_1) A_0(\ell_2) A_0(\ell_3) A_0(\ell_4) + \mbox{3 other terms} \nonumber\\
 &+& 4 \nus \frac{\ell_4}{3 \ell_1 \ell_2 \ell_3} \;  A_1(\ell_1) A_0(\ell_2) A_0(\ell_3) A_1(\ell_4) \; \int_{0}^{\infty} \dd z \; z^3 \; {\rm e}^{-\ell_1 z} K_1(\ell_2 z) K_1(\ell_3 z) K_1(\ell_4 z) + \mbox{11 other terms}  \nonumber\\
&+& \dots
\eea
and
\bea
^{\rm line}t^{\ell_1 \ell_2}_{\ell_3 \ell_4}(L) &=&  \frac{8 \nul }{\ell_1^2 \ell_3^2 L^2} \; A_0(\ell_1) A_0(\ell_2) A_0(\ell_3) A_0(\ell_4) \nonumber\\
&+& 4 \nul \; \frac{\ell_4}{\ell_1^2 \ell_3 L}  A_0(\ell_1) A_0(\ell_2) A_1(\ell_3) A_1(\ell_4) \; \int_{0}^{\infty} \dd z \; z^2 \; {\rm e}^{-\ell_3 z} K_1(\ell_4 z) K_1(L z)  \nonumber\\
 &+& \frac{8 \nul}{\pi^2} \frac{\ell_2 \ell_4}{\ell_1 \ell_3 L} \; A_0(\ell_1) A_1(\ell_2) A_0(\ell_3) A_1(\ell_4) \; \nonumber\\
 &&  \times \int_{-\infty}^{\infty} \dd x \; x  K_1(\ell_1 x) K_1(\ell_2 x) \int_{-\infty}^{\infty} \dd y \; y K_1(\ell_3 y) K_1(\ell_4 y) \;  |x-y|  K_1(L |x-y|)   \nonumber\\
 &+& \left( 1 \leftrightarrow 2 \right) + \left( 3 \leftrightarrow 4 \right) + \left( \begin{array}{crcl} 1 & \leftrightarrow & 2 \\ 3 & \leftrightarrow & 4 \end{array} \right) + \dots \; .
\eea  
For a losange configuration $\ell_1=\ell_2=\ell_3=\ell_4=\ell$, it may be checked that the signs of trispectra do not change since the dominant terms read
\be
^{\rm star}t^{\ell \ell}_{\ell \ell}(L) \simeq \frac{\nus}{ \ell^6} \; A_0^2(\ell) \left[ \frac{32}{3} \, A_0^2(\ell) + \, 7.10 \, A_1^2(\ell) \right] \ee
and
\be
^{\rm line}t^{\ell \ell}_{\ell \ell}(L) \simeq \frac{\nul}{\ell^6}  A_0^2(\ell) \left[ 32 \frac{\ell^2}{L^2} A_0^2(\ell) + 16  \frac{\ell}{L} b_1(1,1,\frac{L}{\ell})  A_1^2(\ell) \right] \; .
\ee
In a similar way, we could expect that the configurations such as $\ell_1=\ell_2=\ell$, $\ell_3=\ell_4=\ell'$ have a constant sign for not too large multipoles $\ell$ and $\ell'$. This seems to be the case in Figs.\ref{fig:TriEqui},~\ref{fig:Tri200} and~\ref{fig:Tri300}. Fig.~\ref{tstar300Approx} shows the reconstructed trispectrum ${\tilde t}^{\ell \; \ell}_{\ell+300 \, \ell+300}(L=5)$. We can check that the trispectrum does not vanish and that its main features are properly reproduced by our expansion in terms of momenta of the gravitational field.

\section{Conclusions}

We investigated the shapes of high order correlation functions in the CMB anisotropy maps. Formal expressions 
for the high order correlation functions can easily be obtained. We found that, for generic models of primordial non-Gaussianities, they can be described with the help of formal diagrams evaluated with computation rules we gave. We found that only three kinds of functions with simple diagrammatic interpretations are needed: the radial transfer functions $R_{\ell}(r)$ represented by vertices, the propagators $\xi_{\ell}(r)$ represented by outer lines joining a point to a vertex and the vertex propagators, $\zeta_L(r;r')$, corresponding to internal lines joining two vertices. 
This formalism provides simple expressions of high order connected correlation functions of the observed temperature field for a large class of models of primordial non-Gaussianities. 

The development of instabilities at super-horizon scales is intrinsically a nonlinear process that induces specific non-Gaussian effects. These types of mode couplings naturally induce a potential bispectrum the amplitude of which is characterized by a parameter $\nu_{2}$ (e.g. $f_{NL}$) of the order unity. However for such a low level of couplings and beyond the Sachs Wolfe regime, the calculation of the temperature correlation functions requires a consistent second order treatment of the CMB anisotropies through the physics of recombination which is not taken into account in the formalism we present.

More significant sources of non-Gaussianities might come from additional scalar degrees of freedom in the very early universe. In such cases, the statistics induced may depend on the details of the scenario and in particular on the types of mode couplings during or at the end of the inflationary period. It makes the search of non-Gaussianities of the observed CMB maps precious for discriminating between different inflationary scenarii. This is illustrated for instance by the qualitative differences we found at the level of the tri-spectra between {\em line}- and {\em star}-like mode coupling effects. Those shapes are furthermore found to be quite different from those induced by weak lensing effects.  

As we stressed it is possible to obtain formally compact expressions for the high order correlation functions. It is nonetheless difficult
to get insights into the geometrical dependences of these correlation functions in general. Projection effects and physics of recombination both contribute to make difficult the transcription from the high order correlation functions of the gravitational potential to those of the observed temperature field. Only in the Sachs Wolfe regime is this transcription easy. The same formal structure between the correlation functions is indeed recovered. To obtain further insights into the behaviors of the high order correlation functions at small angular scale, we propose a description of the acoustic oscillations using semi-analytic methods. Ignoring the integrated Sachs-Wolfe effect, we found that it was possible to use a gradient-like expansion as suggested by the fact that the radial transfer function is peaked in the vicinity of the last scattering surface. This approximation assumes that the temperature anisotropies can only be described by the behavior of the potential at or in the vicinity of the last scattering surface.
In the approach we adopted the transfer physics is encoded by angular scale dependent terms while projection effects are treated separately. In this perspective, the small angle approximation is appropriate to handle projection effects and leads to more convenient results. We found that the shapes of the temperature power spectrum and bispectrum can be reasonably reproduced with the very first terms of such an expansion. 

This perturbative approach can give insights into the behavior of the high order correlation functions. For instance it gives a good account of the positions of the zeros of the bispectrum. They indeed have to roughly coincide with those of the monopole term at the surface of last scatter. Higher order expansion obviously leads to an increasing accuracy in the description of the amplitude of the temperature bispectrum. The trispectra show more complex structures with complicated sign patterns but that can also be explained, to a large extent, with the approach we have developed. In particular, some key features of the trispectrum, such as its sign in an equilateral configuration, may be accounted for. Nevertheless, the features of the high order temperature spectra might be complementary highlighted in a real space treatment of the CMB anisotropies~\cite{CMBreal}. We think that the method we have presented could be extended to the properties of the polarization field as well.

\begin{acknowledgements}

We are grateful to Jean-Philippe Uzan whom the project was initiated with. We also thank Alain Riazuelo for fruitful discussions.

\end{acknowledgements}



\appendix 
 
\section{Functions involved in the expansion of the transfer function} \label{AppExpansion}

Expanding the transfer function as in Eq.~(\ref{Texpand}) induces the calculation of specific integrals in the expression of the radial transfer function $R_\ell(r)$ or in the function $\xi_\ell(r)$. In this subsection, we will give the expressions of the function $g_i(r, \rs;\ell)$ and $f_i(r/\rs;\ell)$ (see Eqs.~(\ref{Rexpand})-(\ref{fn})). 

To compute these integrals, we will use the following equality (see \cite{gradshteyn}) for $-2<p<\ell+\ell'+1$ and $r<\rs$: 
\bea
\label{intbes1}
\int \dd k \; k^{-p} j_{\ell'}(k \rs) j_{\ell}(k r) &=& \frac{\pi}{2^{p+2}} \frac{r^{\ell}}{\rs^{\ell-p+1}} \frac{\Gamma(\frac{\ell+\ell'-p+1}{2})}{\Gamma(\frac{\ell'-\ell+p+2}{2}) \Gamma(\ell+3/2)} \times \nonumber \\
&& _2 \! F_1 \left( \frac{\ell+\ell'-p+1}{2},\frac{\ell-\ell'-p}{2} ; \ell+3/2;\frac{r^2}{\rs^2}\right) \;, 
\eea
whereas for $r>\rs$
\bea
\label{intbes2}
\int \dd k \; k^{-p} j_{\ell'}(k \rs) j_{\ell}(k r) &=& \frac{\pi}{2^{p+2}} \frac{\rs^{\ell'}}{r^{\ell'-p+1}} \frac{\Gamma(\frac{\ell+\ell'-p+1}{2})}{\Gamma(\frac{\ell-\ell'+p+2}{2}) \Gamma(\ell'+3/2)}  \times \nonumber \\
&& _2 \! F_1 \left( \frac{\ell+\ell'-p+1}{2},\frac{\ell'-\ell-p}{2} ; \ell'+3/2;\frac{\rs^2}{r^2}\right) \; .
\eea
On the other hand, the recurrence relation between spherical Bessel functions
\be
\label{recbes}
j'_\ell(x) = \frac{1}{2 \ell +1} \left[ \ell j_{\ell-1}(x) - (\ell+1) j_{\ell+1}(x)\right]\;, 
\ee
leads to
\be
\label{recbesgen}
j_{\ell}^{(p)}(x) = \sum_{\ell'=\ell-p}^{\ell+p} \alpha_{\ell \ell'}^{(p)} j_{\ell'}(x) \;, 
\ee
where the coefficients $\alpha_{\ell, \ell'}^{(p)}$ satisfy the recurrence relation
\be
\label{coeffrec}
\alpha_{\ell, \ell'}^{(p+1)} = \frac{\ell'+1}{2\ell'+3} \; \alpha_{\ell, \ell'+1}^{(p)}-\frac{\ell'}{2\ell'-1} \; \alpha_{\ell, \ell'-1}^{(p)}\; .
\ee

The integral defining $g_1$ in Eq.~(\ref{gn}) may be computed with Eqs.~(\ref{intbes1})-(\ref{recbes}) assuming that the limit $p=-2$ may be analytically continued. The expression for $g_1$ yields for $r<\rs$
\be
g_1(r,\rs;\ell) = \frac{\sqrt{\pi}}{2 \rs^2} \frac{\Gamma(\ell +1)}{\Gamma(\ell+3/2)} \left( \frac{r}{\rs}\right)^{\ell} \; \left[ -\frac{\ell+2}{\rs} \;  {_2} \! F_1 \left( \frac{1}{2},\ell+1,\ell+\frac{3}{2};\frac{r^2}{\rs^2} \right) - \frac{\ell+1}{\ell+3/2} \; \frac{1}{\rs^3}\; {_2} \! F_1 \left( \frac{3}{2},\ell+2,\ell+\frac{5}{2};\frac{r^2}{\rs^2} \right) \right]\; .
\ee
Similarly for $r>\rs$
\be
g_1(r,\rs;\ell) = \frac{\sqrt{\pi}}{2 r^2} \frac{\Gamma(\ell +1)}{\Gamma(\ell+3/2)} \left( \frac{\rs}{r}\right)^{\ell} \; \left[ -\frac{\ell}{\rs} \;  {_2} \! F_1 \left( \frac{1}{2},\ell+1,\ell+\frac{3}{2};\frac{\rs^2}{r^2} \right) + \frac{\ell}{\ell+3/2} \; \frac{\rs}{r^2}\; {_2} \! F_1 \left( \frac{3}{2},\ell+2,\ell+\frac{5}{2};\frac{\rs^2}{r^2} \right) \right]\; .
\ee
Here $_2 \! F_1$ stands for the hypergeometric function.
In the computation of $g_2$, the integrals in Eq.~(\ref{intbes1}) are not defined for $p=-2$. Hence, we should write $g_2$ as
\be
g_2(r,\rs;\ell) =  \frac{\dd^2}{\dd \rs^2} \int \dd k \; j_{\ell}(k r) j_{\ell}(k \rs) \; .
\ee

Using Eqs.~(\ref{intbes1}) and (\ref{intbes2}), we get
\be
 \int \dd k \; j_{\ell}(k r) j_{\ell}(k \rs) = \left\{ \begin{array}{c}  \frac{\pi}{2} \; \frac{1}{2 \ell + 1} \; \frac{1}{\rs} \left(\frac{r}{\rs}\right)^{\ell} \; \; \mbox{for $r<\rs$} \\ \frac{\pi}{2} \; \frac{1}{2 \ell + 1} \; \frac{1}{r} \left(\frac{\rs}{r}\right)^{\ell} \; \; \mbox{for $r>\rs$} \end{array} \right. \;   
\ee
and the function $g_2$ reads
\be
g_2(r,\rs;\ell) =  \left\{ \begin{array}{c}  \frac{\pi}{2} \; \frac{(\ell+1)(\ell+2)}{2 \ell + 1} \; \frac{1}{\rs^3} \left(\frac{r}{\rs}\right)^{\ell} \; \; \mbox{for $r<\rs$} \\ \frac{\pi}{2} \; \frac{\ell(\ell-1)}{2 \ell + 1} \; \frac{1}{\rs^3} \left(\frac{\rs}{r}\right)^{\ell+1} \; \; \mbox{for $r>\rs$} \end{array} \right. \; .
\ee

The expressions of the integrals $f_i$ in Eq.~(\ref{fn}) may be performed with Eqs.~(\ref{intbes1})-(\ref{coeffrec}) and yield for $r<\rs$
\bea 
f_0(\frac{r}{\rs};\ell) &=& \frac{\sqrt{\pi}}{4} \frac{\Gamma(\ell)}{\Gamma(\ell+3/2)} \left( \frac{r}{\rs}\right)^{\ell} \; _2 \! F_1 (\ell,-\frac{1}{2};\ell+\frac{3}{2};\frac{r^2}{\rs^2})  \\ 
f_1(\frac{r}{\rs};\ell) & = &  \frac{\pi}{4} \left(\frac{r}{\rs}\right)^{\ell} \left[ -\frac{ \ell -1 }{(2 \ell -1 )(2 \ell +1 )} + \frac{ \ell +1 }{(2 \ell +1 )(2 \ell +3 )} \left( \frac{r}{\rs} \right)^{2} \right] \\ 
f_2(\frac{r}{\rs};\ell) & = & \frac{\sqrt{\pi}}{8} \left(\frac{r}{\rs}\right)^{\ell} \frac{\Gamma(\ell+1)}{\Gamma(\ell+3/2)} \times  \nonumber \\ 
&& \Bigg[ \frac{4(\ell+1)(\ell+2)}{3(2\ell+1)(2\ell+3)} \;  _2 \! F_1 (\ell+1,-\frac{3}{2};\ell+\frac{3}{2};\frac{r^2}{\rs^2}) - 2\frac{2\ell^2+2\ell-1}{\ell (2\ell-1)(2\ell+3)} \;  _2 \! F_1 (\ell,-\frac{1}{2};\ell+\frac{3}{2};\frac{r^2}{\rs^2})   \nonumber \\ 
&&+ \frac{1}{(2\ell-1)(2\ell+1)} \;  _2 \! F_1 (\ell-1,\frac{1}{2};\ell+\frac{3}{2};\frac{r^2}{\rs^2}) \Bigg] \; . 
\eea
In the case $r>\rs$, they transform into
\bea
f_0(\frac{r}{\rs};\ell) &=& \frac{\sqrt{\pi}}{4} \frac{\Gamma(\ell)}{\Gamma(\ell+3/2)} \left( \frac{r}{\rs}\right)^{\ell} \; _2 \! F_1 (\ell,-\frac{1}{2};\ell+\frac{3}{2};\frac{\rs^2}{r^2}) \\ 
f_1(\frac{r}{\rs};\ell) & = &  \frac{\pi}{4} \left(\frac{\rs}{r}\right)^{\ell} \left[ \frac{ \ell  }{(2 \ell -1 )(2 \ell +1 )} - \frac{ \ell +2 }{(2 \ell +1 )(2 \ell +3 )} \left( \frac{\rs}{r} \right)^{2} \right] \\
f_2(\frac{r}{\rs};\ell) & = & \frac{\sqrt{\pi}}{8} \left(\frac{\rs}{r}\right)^{\ell-2} \frac{\Gamma(\ell+1)}{\Gamma(\ell+3/2)} \times  \nonumber \\ 
&& \Bigg[ \frac{1}{3} \;  _2 \! F_1 (\ell-1,-\frac{3}{2};\ell-\frac{1}{2};\frac{\rs^2}{r^2}) - 2\frac{2\ell^2+2\ell-1}{\ell (2\ell-1)(2\ell+3)} \left(\frac{\rs}{r}\right)^2 \;  _2 \! F_1 (\ell,-\frac{1}{2};\ell+\frac{3}{2};\frac{\rs^2}{r^2})   \nonumber \\ 
&&+ \frac{4(\ell+1)(\ell+2)}{(2\ell+1)(2\ell+3)^2(2\ell+5)} \left(\frac{\rs}{r}\right)^4 \;  _2 \! F_1 (\ell+1,\frac{1}{2};\ell+\frac{7}{2};\frac{\rs^2}{r^2}) \Bigg] \; . 
\eea

\section{Integrals involved in the computation of the bispectrum} \label{Appbis}
 
The computation of the bispectrum involves integrals such as
\be
\int_{-\infty}^{+\infty} \dd x \; \frac{x^p}{(x^2+\ell^2)^{(p+3)/2}} {\rm e}^{-i x z} \; .
\ee

The integrals for $p=0$, 1 and 2 may be performed~\cite{gradshteyn}
\bea
\label{K1}
\int_{-\infty}^{\infty} \dd x \; \frac{{\rm e}^{-i x z}}{(\ell^2+x^2)^{3/2}} &=& \frac{2}{\ell} |z| K_1(\ell |z|)\;, \\
\int_{-\infty}^{\infty} \dd x \; \frac{x \; {\rm e}^{-i x z}}{(\ell^2+x^2)^{2}} &=& -i \frac{\pi}{2\ell} z {\rm e}^{-\ell |z|}\;, \\
\int_{-\infty}^{\infty} \dd x \; \frac{x^2 \, {\rm e}^{-i x z}}{(\ell^2+x^2)^{5/2}} &=& \frac{4}{3\ell^2} G_{13}^{21} \left( \frac{\ell^2 z^2}{4} \Big | \begin{array}{cc} - 1 / 2 &\\ 0 1 & 1 / 2 \end{array} \right) \;, 
\label{meijer}
\eea
where
$K_n$ is the modified Bessel function of the second kind of order $n$ and $G^{m, n}_{p, q} \left( x \Big | \begin{array}{cc}  a_1,...,a_p \\ b_1,...,b_q \end{array} \right)$ is the Meijer's G-function. The Meijer's G-function is defined by
\be
G^{m, n}_{p, q} \left( x \Big | \begin{array}{cc}  a_1,...,a_p \\ b_1,...,b_q \end{array} \right) = \frac{1}{2 i \pi} \int_{{\cal C}} \frac{\prod_{j=1}^{m} \Gamma(b_j+s) \prod_{j=1}^{n} \Gamma(1-a_j-s)}{\prod_{j=m+1}^{q} \Gamma(1-b_j-s) \prod_{j=n+1}^{p} \Gamma(a_j+s)} x^{-s} {\rm d}s \;, 
\ee
where the contour ${\cal C}$ lies between the poles of $\Gamma(1-a_j-s)$ and the poles of $\Gamma(b_j+s)$.
The last integral~(\ref{meijer}) may be simplified using
\be
\int_{-\infty}^{\infty} \dd x \; \frac{x^2 \, {\rm e}^{-i x z}}{(\ell^2+x^2)^{5/2}} = \frac{1}{3 \ell} \frac{\partial^2}{\partial z^2} \, \frac{\partial}{\partial \ell} \int_{-\infty}^{\infty} \dd x \; \frac{{\rm e}^{-i x z}}{(\ell^2+x^2)^{3/2}}\; .
\ee
With Eq.~(\ref{K1}) and the differential equations satisfied by the Bessel functions, we obtain
\be
\int_{-\infty}^{\infty} \dd x \; \frac{x^2 \, {\rm e}^{-i x z}}{(\ell^2+x^2)^{5/2}}=\frac{2}{3 \ell^2} \left[ \ell |z| K_1(\ell |z|)-\ell^2 z^2 K_0(\ell |z|) \right]\; .
\ee

Some other integrals involved in the computation of the bispectrum are of the type
\be
\int_{-\infty}^{\infty} \dd x \; \frac{x^p}{(x^2+\ell^2)^{p/2}} {\rm e}^{-i x z} \; .
\ee
Results for $p=0$, 1 and 2 read
\bea
\int_{-\infty}^{\infty} \dd x \;  {\rm e}^{-i x z} &=& 2 \pi \delta_{\rm Dirac}(z) \;, \\
\int_{-\infty}^{\infty} \dd x \; \frac{x \; {\rm e}^{-i x z}}{(\ell^2+x^2)^{1/2}} &=& 2 -i \, \ell \; {\rm sgn(z)} \;  K_1(\ell |z|) \;, \\
\int_{-\infty}^{\infty} \dd x \; \frac{x^2 \, {\rm e}^{-i x z}}{\ell^2+x^2} &=& - \pi \; \ell \; {\rm e}^{-\ell |z|} \; .
\eea
For the special cases $z=0$ 
\bea 
\int \dd x \; \frac{1 }{(\ell^2+x^2)^{3/2}} &=& \frac{2}{\ell^2}\;, \\ 
\int \dd x \; \frac{x}{(\ell^2+x^2)^2} &=& 0 \;, \\ 
\int \dd x \; \frac{x^2}{(\ell^2+x^2)^{5/2}} &=& \frac{2}{3\ell^2} \; . 
\eea 

\section{Flat sky formalism}
\label{sec:flatsky}

We define the quantity $a(\vl)$ by the bidimensional Fourier transform of the temperature contrast
\be
\label{avecl}
a(\vl) = \int \frac{\dd^2 \hg}{2\pi} \; \frac{\delta T}{T}(\hg) {\rm e}^{-i \vl \cdot \hg}
\ee
or conversely
\be
\frac{\delta T}{T}(\hg) =  \int \frac{\dd^2 \vl}{2\pi} \; a(\vl) {\rm e}^{i \vl \cdot \hg}.
\ee
Inserting the multipole expansion of the temperature contrast in Eq.~\ref{avecl}, we may express $a(\vl)$ as a function of the coefficients $a_{\ell' m'}$
\be
a(\vl) = \sum_{\ell' m'} a_{\ell' m'} \int \frac{\dd^2 \hg}{2\pi} \; Y_{\ell' m'}(\hg) \, {\rm e}^{-i \vl \cdot \hg}.
\ee
In the small angle approximation, the spherical harmonics are properly approximated by
\be
Y_{\ell m}(\theta,\phi) \simeq J_m(\ell \theta) \sqrt{\frac{\ell}{2\pi}} {\rm e}^{i m \phi} \; 
\ee
for $\ell \gg 1$ and $\theta \ll 1$.
Moreover, we can decompose the vectors $\hg$ and $\vl$ in the small angle approximation as
\be
\hg \left( \begin{array}{c} \theta \cos \phi \\ \theta \sin \phi \end{array}\right)  \; \; \mbox{and} \; \; \vl\left( \begin{array}{c} \ell \cos \phi_\ell \\ \ell \sin \phi_{\ell} \end{array}\right). 
\ee
The quantity $a(\vl)$ then becomes
\be
a(\vl) = \sum_{\ell' m} a_{\ell' m} \int \frac{\dd^2 \hg}{2 \pi} \;  J_m(\ell' \theta) \sqrt{\frac{\ell'}{2\pi}} {\rm e}^{i m \phi} {\rm e}^{-i \ell \theta \cos(\phi-\phi_{\ell})}. 
\ee
Using the integral representation of the Bessel functions
\be
\label{Jmrep}
J_m(\ell \theta) = \frac{1}{2 \pi i^m} \int_0^{2\pi} \dd \phi' \; {\rm e}^{i m \phi'+i \ell \theta \cos \phi'},
\ee
one naturally gets
\be
a(\vl) = \sqrt{2\pi\, \ell'} \; \sum_{\ell' m}  (-i)^{m} \, a_{\ell' m} \, {\rm e}^{-i m \phi_\ell} \int \frac{\dd \theta}{2\pi} \; \theta J_{m}(\ell \theta) J_{m}(\ell' \theta),
\ee
which leads to
\be
\label{avecl2}
a(\vl) = \frac{1}{\sqrt{2\pi \, \ell}} \sum_m (-i)^m \,  a_{\ell m} \, {\rm e}^{-i m \phi_\ell}.
\ee
Conversely
\be
a_{\ell m} = \sqrt{\frac{\ell}{2\pi}} i^m \int \dd \phi_{\ell} \; a(\vl) {\rm e}^{i m \phi_\ell}.
\ee

Defining the power spectrum as
\be
\langle a_{\ell m} a_{\ell' m'}^* \rangle =  C_{\ell} \; \delta_{\ell \ell'} \delta_{m m'},
\ee
we find, using Eq.~(\ref{avecl2}) that
\be 
\langle a(\vl) a(\vl') \rangle = \delta_{\rm Dirac}(\vl-\vl') \; C_{\ell}.
\ee

In an all sky formalism, the reduced bispectrum is defined by
\be
\langle a_{\ell_1 m_1} a_{\ell_2 m_2}  a_{\ell_3 m_3} \rangle_c = {\cal G}^{m_1 m_2 m_3}_{\ell_1 \ell_2 \ell_3} b_{\ell_1 \ell_2 \ell_3} \;. 
\ee
Using Eqs.~(\ref{avecl2}) and~(\ref{Jmrep}), we find the power spectrum in a flat sky formalism
\be
\langle a(\vl_1) a(\vl_2)  a(\vl_3) \rangle_c = \frac{1}{2\pi} \;  b_{\ell_1 \ell_2 \ell_3} \; \delta_{\rm Dirac}(\vl_1+\vl_2+\vl_3) \; .
\ee

The reduced trispectrum is defined by
\be
\langle a_{\ell_1 m_1} a_{\ell_2 m_2}  a_{\ell_3 m_3}  a_{\ell_4 m_4} \rangle_c = \sum_{L,M} (-1)^M {\cal G}_{\ell_1 \ell_2 L}^{m_1 m_2 M} {\cal G}_{\ell_3 \ell_4 L}^{m_1 m_2 -M} t^{\ell_1 \ell_2}_{\ell_3 \ell_4}(L)
\ee 
and the correspondence between flat sky and full sky formalisms imposes
\be
\langle a(\vl_1) a(\vl_2) a(\vl_3) a(\vl_4) \rangle_c =   \frac{1}{(2\pi)^2} \int \frac{\dd^2L}{2\pi} \; \delta_{\rm Dirac}(\vl_1+\vl_2+\vl_3+\vl_4) \; \delta_{\rm Dirac}(\vl_1+\vl_3-\vec{L}) \; t^{\ell_1 \ell_2}_{\ell_3 \ell_4}(L)\;. 
\ee

\end{document}